\documentclass[12pt,preprint]{aastex}

\shorttitle{ISO photometry of 12$\mu$m active galaxies}
\shortauthors{Spinoglio et al.}

\begin{document}


\title{The far-infrared energy distributions of Seyfert and starburst
galaxies in the Local Universe: ISO photometry of the 12$\mu$m
active galaxy sample\footnote{Based on observations with ISO, an
ESA project with instruments funded by ESA Member States
(especially the PI countries: France, Germany, the Netherlands
and the United Kingdom) with the participation of ISAS and NASA}}


\author{Luigi Spinoglio}\affil{Istituto di Fisica dello Spazio Interplanetario, CNR
via Fosso del Cavaliere 100, I-00133 - Roma (Italy)}
\email{luigi@ifsi.rm.cnr.it}

\author{Paola Andreani\altaffilmark{2}}\affil{Osservatorio Astronomico Padova,
Vicolo dell'Osservatorio 5, I-35122 - Padova (Italy)}
\email{andreani@pd.astro.it}

 \and

\author{Matthew A. Malkan}\affil{Physics and Astronomy Department, University of California
at Los Angeles, Los Angeles, CA 90095-1562 (USA)}
\email{malkan@astro.ucla.edu}

\altaffiltext{2}{Present address: Max-Planck-I. f\"{u}r
Extraterrestrische Physik, Postfach 1312, 85741 Garching (Germany)
}


\begin{abstract}
New far-infrared photometry with ISOPHOT, onboard the {\it
Infrared Space Observatory}, is presented for 58 galaxies with
homogeneous published data for another 32 galaxies all belonging
to the 12${\mu}m$ galaxy sample. In total 29 Seyfert~1's, 35
Seyfert~2's and 12 starburst galaxies, about half of the 12$\mu$m
active galaxy sample, plus 14 normal galaxies for comparison. The
ISO and the IRAS data are used to define color-color diagrams and
spectral energy distributions (SED). Thermal dust emission at two
temperatures (one cold at 15-30K and one warm at 50-70K) can fit
the 60--200${\mu}m$ SED, with a dust emissivity law proportional
to the inverse square of the wavelength. Seyfert 1's and Seyfert
2's are indistinguishable longward of 100$\mu$m, while, as
already seen by IRAS, the former have flatter SEDs shortward of
60$\mu$m. A mild anti-correlation is found between the [200 - 100]
color and the ``60$\mu$m excess". We infer that this is due to
the fact that galaxies with a strong starburst component, and thus
a strong 60$\mu$m flux, have a steeper far-infrared turnover.  In
non-Seyfert galaxies, increasing the luminosity corresponds to
increasing the star formation rate, that enhances the 25 and
60$\mu$m emission. This shifts the peak emission from around
150$\mu$m in the most quiescent spirals to shorter than 60$\mu$m
in the strongest starburst galaxies. To further quantify these
trends, we followed \citet{RR89} in identifying with the IRAS
colors three idealized infrared SED: that of pure quiescent disk
emission, pure starburst emission, and pure Seyfert nucleus
emission. Even between 100 and 200$\mu$m, the quiescent disk
emission remains much cooler than the starburst component.
Seyfert galaxies have 100--200$\mu$m SED ranging from pure disks
to pure starbursts, with no apparent contribution from their
active nuclei at those wavelengths.

\end{abstract}



\section{INTRODUCTION}

One of the main goals of selecting a complete sample of galaxies
using the IRAS 12${\mu}m$ flux was the definition of a complete
and largely unbiased sample of active galaxies in the local
universe. This selection was done twice, the first one producing
a list of 26 Seyfert type 1 galaxies (hereinafter Seyfert 1's) and
32 Seyfert 2's \citep{spi89}, out of a sample of 390 galaxies
from the \citet{iras88}. The second was a selection of 53 Seyfert
1's, 63 Seyfert 2's, two blazars and 38 high-luminosity
non-Seyferts (i.e. galaxies with $L_{IR} \geq 10^{44} {\rm
erg~s^{-1}}$ without an obvious Seyfert type optical spectrum) out
of a sample of 893 galaxies \citep{RMS93} (RMS) from the IRAS
Faint Source Catalog \citep{mo92}. It was found that the
12${\mu}m$ flux is approximately a constant fraction ($\sim$ 1/5)
of the bolometric flux in active galaxies. Moreover, also for
non-Seyferts (mostly spirals) 12${\mu}m$ selection was found to be
the closest available approximation to selection by a limiting
bolometric flux, which is about 14 times $\nu~F_{\nu}$ at
12${\mu}m$ \citep{S95} (hereinafter S95). It therefore follows
that deep surveys at 12${\mu}m$ should provide complete samples at
different bolometric flux levels of normal and active galaxies,
which will not suffer the strong selection effects present both
in the optical-UV and far-infrared. We refer to the introduction
of the paper presenting the original sample \citep{spi89} for the
arguments in favor of the mid-infrared selection. After the
determination by RMS, two more works derived the local 12${\mu}m$
luminosity function \citep{XU98,FA98} and a new selection of
galaxies from the IRAS database at 12${\mu}m$ is in progress
\citep{A99}.

The study of the infrared energy distributions of active and
starburst galaxies in the local universe is not only important by
itself, but also because it is needed to understand galaxy
evolution. Deep ISO surveys have shown that indeed a strong
evolution is present at z $\sim$ 1, indicating that many galaxies
at that epoch were experiencing phases of extremely high
luminosities in the far-infrared, likely to represent violent
events of star formation [see e.g. the reviews of \citet{MA00},
\citet{GC00}, and \citet{FRA01}]. Also, the peak of the energy
output of galaxies lies in the far-infrared. This makes
redshifted galaxies appear relatively brighter at the longest
far-infrared and sub-millimeter wavelengths; surveys in these
wavebands are therefore capable of discovering very distant
galaxies at substantial cosmic look-back times \citep[see
e.g.][]{MA01}. It is thus even more urgent to understand the local
infrared emission, to provide a firm base for comparing the local
universe ``activity" with the properties and evolution observed in
the recent and upcoming cosmological surveys
\citep[e.g.][]{FRA01,MA01}.

Several papers have reported ISO photometric data on galaxies in
the unexplored range from 100-200${\mu}m$, but most of these are
on few individual sources, often the prototypes of the various
classes of active, starburst and normal galaxies. The only
exceptions are the study of 42 objects (20 Seyfert 1's and 22
Seyfert 2's) \citep{P-G00}(PGRE) of the CfA Seyfert galaxy sample
\citep{HB92}, the study of a random sample \citep{H00} of 17
Palomar Green quasars \citep{SG83}, the study of 22 radio-loud
and radio-quiet quasars \citep{PO00} and one of 10 galaxy/quasar
pairs from the 3CR catalog \citep{mei01}. The CfA Seyfert galaxy
sample is smaller (48 objects) than the 12${\mu}m$ active galaxy
sample (158 objects), and is selected by the optical light of the
galaxy, which may bias the results significantly. For example, a
recent work on high resolution radio observations of the Seyfert
galaxies of the 12${\mu}m$ sample \citep{th00,th01}, report a
detailed statistical analysis of the redshift distributions and
12${\mu}m$ flux distributions of Seyfert 1's and 2's and compared
these with the CfA sample. In the 12${\mu}m$ active galaxy sample
the two types of Seyferts are equally distributed according to
redshift, their 12${\mu}m$ flux density distributions are well
matched and therefore Seyfert 1's and 2's are equally luminous in
the IRAS 12${\mu}m$ band. On the contrary, the CfA Seyfert sample
has different redshift distributions for the two Seyfert types,
and is biased against including distant Seyfert 2's.

The three latter works on quasars \citep{H00,PO00,mei01} are
complementary to ours since most of the sources have much higher
redshifts (${\rm 0.1<z<2.0}$). However, these studies are not
suited to studying the population of active galaxies in the local
universe.

The present work is based on observations of 90 galaxies,
randomly selected from the 12${\mu}m$ galaxy sample, including 29
Seyfert 1's, 35 Seyfert 2's, 12 high luminosity non-Seyferts
(hereafter called starburst galaxies) and 14 normal galaxies for
comparison. It is the largest sample of active galaxies for which
far-infrared photometry is available out to 200${\mu}m$. Since it
contains both active and starburst galaxies it is expected to be
the basis for firm and statistically significant conclusions on
the far-infrared behaviour of galaxies in the local universe.

The paper is organized as follows: section 2 describes the
observations collected and reduced longward of 100${\mu}m$;
section 3 shows the results in terms of average energy
distributions and presents a derivation of ``pure" infrared
emission components, for quiescent disks, starbursts, and Seyfert
nuclei; section 4 shows the color-color diagrams of the various
classes of galaxies that can be explained by thermal emission
from dust and presents a correlation between the far infrared
color [200 - 100] and the 60${\mu}m$ excess; section 5 shows the
correlations among the different luminosities and between colors
and luminosities. Section 6 presents our conclusions. Finally, in
the Appendix we present the results obtained shortward of
100${\mu}m$.

\section{OBSERVATIONS}

The Infrared Space Observatory (ISO) \citep{K96} data presented
here were obtained in the open time proposal ``IR energy
distributions and imaging of the complete sample of 12${\mu}m$
active galaxies". Photometric data at several wavelengths between
60 and 200 $\mu m$ were collected with the two array cameras,
C100 and C200 of the ISOPHOT imaging photo-polarimeter
\citep{L96}. In particular, we present in this section the
ISOPHOT photometric results obtained with the C200 detector array
in the spectral range 120-200${\mu}m$ on 39 objects belonging to
the 12${\mu}m$ active galaxies sample. The data obtained with the
C100 array camera are presented in Appendix 1, for two reasons:
first, these data are less precise than the C200 data, having
larger uncertainties, due either to a difficult subtraction of
the background and to a less accurate calibration (see the
discussion in Appendix 1); second, these data are less important,
because all the galaxies of the sample have good IRAS detections
at 12, 25, 60 and 100$\mu$m. To increase the statistics, we have
also analyzed ISOPHOT C200 archive data on other 19 galaxies and
we have included in the discussion ISOPHOT literature data on
another 32 galaxies, bringing the present sample to 90 objects.
This is therefore currently the largest sample of active galaxies
in the local universe studied to 200$\mu$m.

The journal of the C200 ISOPHOT observations is given in Table 1,
which presents the details of each single observation that we
analysed: for each source are given the coordinates of the
observations, the redshift, the galaxy type (hereafter we call
starburst nuclei {\it sbn} the high luminosity non Seyfert
galaxies), the ISO observation identification number and the total
integration time spent on-source.

All of the detections are consistent with that expected for
unresolved point sources. That is, none of the galaxies were
significantly resolved by the large ISOPHOT PSF, which is
estimated to be $\sim$ 100 $\arcsec$ FWHM in diameter at 120
$\mu$m [e.g. see Fig.4.8 of \citet{Lau01}].

The C200 ISOPHOT observations were collected in chopped
photometric mode in the five filters centered at 120, 150, 170,
180 and 200${\mu}m$. The data reduction has been performed using
the PIA\footnote{PIA is a joint development by the ESA
Astrophysics Division and the ISOPHOT consortium led by the MPI
fur Astronomie, Heidelberg. Contributing institutes are DIAS, RAL,
AIP, MPIK and MPIA.} software, version 7.2 \citep{G97} and the
results were verified with version 8.0. Data were deglitched,
corrected for non-linearity and then for each ramp a straight
line was fitted. In most cases the first signal per chopper
plateau was discarded since the detector response was not
stabilized. Then the data were corrected for dark current and for
signal dependence on the ramp integration time and calibrated
with the internal Fine Calibration Source (FCS1). The same
reduction procedure was applied also to the calibration data. An
independent reduction of several of these observations was made
by one of us (MM) at IPAC\footnote{IPAC (Infrared Processing and
Analysis Center) is part of the Division of Physics, Mathematics
and Astronomy at Caltech, Pasadena, and the Space and Earth
Sciences Program Directorate at JPL, Pasadena.}, those results are
in good agreement with the values published here. We realize there
are faint outer parts to some of these galaxies, and this was even
measured and discussed in the appendix of RMS. However, we find
that only a small fraction of the total flux is actually missed
due to the central concentration of the light, and that our
colors should be very reliable anyway, one of the main points of
this study.


The ISOPHOT flux densities have been corrected using the color
correction, from the PIA \citep{G97}, corresponding to black-body
functions with an emissivity law proportional to the inverse
square of the wavelength. The chosen dust temperatures were the
lowest color temperature that we derive from the [200 - 100]
color of each class of galaxies, derived from the positions in
the far-infrared color-color diagrams presented in section 4. They
are 20K for Seyfert 1's and normal galaxies, 25K for Seyfert 2's
and 30K for high luminosity non-Seyferts. At these temperatures,
the color correction is, however, quite small (less than 15\%).

Because all the galaxies have good detections in all the IRAS
bands, we have constructed their spectral energy distributions
using the new ISOPHOT data together with the IRAS data given in
RMS. In Fig. 1 we present the spectral energy distributions of
each individual object. Table 2 presents the measured flux
densities and 1~$\sigma$ statistical errors in the five ISOPHOT
C200 wavebands for each object. Absolute calibration uncertainty
is in the range 10---30\%.

For most of the sources a good general agreement is found between
IRAS and ISO measurements and the C200 values place themselves on
a smooth curve running from the IRAS to the ISO photometry. Data
for Arp~220, NGC~6240 and IZW~1 were taken from the archive. The
results reported here agree well with those published in the
literature for the first two sources \citep{K97} while the C200
data of IZW~1 of Table 2 agree with those published by \citet{H00}
but not with those by PGRE. The measurements were taken with
different observational setups, the ones reported here were
gathered in chopping mode while those by PGRE in staring mode and
the resulting values are much lower in this latter case. We
ascribe this difference to the different background subtraction
procedure and it could be due to an overestimation of the
background value by this latter authors. They indeed did not
detect IZw~1 at 200$\mu$m while we obtained a signal-to-noise
ratio larger than 18 at the same wavelength. Data taken towards
NGC~6090 (MK~496) reported here are lower than those published by
\citet{CA00} by a factor of 1.3-1.7, however they agree within
the large errors quoted by these authors.

In the following analysis we also included the ISO results
published by several authors on galaxies belonging to the
12$\mu$m galaxy sample: 17 CfA Seyfert galaxies (PGRE), 10 normal
galaxies \citep{S99}, 4 nearby spiral galaxies \citep{A98} and
the Seyfert 2 NGC~7582 \citep{RA99}. The CfA Seyfert 1's included
are: NGC~3227, NGC~3516, NGC~4051, NGC~4151, MK~766, MK~231,
NGC~5033, NGC~5548, MK~817 and NGC~7469, while the CfA Seyfert
2's are: NGC~1143/44, NGC~3079, NGC~3982, NGC~4388, NGC~5256,
NGC~5929 and NGC~1068. The normal galaxies included are: MK~323,
MK~332, MK~555, MK~538, MK~799, NGC~5719, NGC~6918, NGC~7083,
UGC~2936 and UGC~2982 and the nearby spiral galaxies are:
NGC~134, NGC~660, NGC~5194 and NGC~5236.

Table 3 gives the integrated mid-to-far infrared luminosity from
8 to 250$\mu$m for all 90 galaxies, including those taken from the
literature. This luminosity has been computed using the formula:

\begin{equation}
 L_{FIR} = 4 \pi (\frac{cz}{H_{o}})^{2} \times \sum_{i} (S_{\nu_{i}}
\times \Delta\nu_{i}) \times 10^{-11}{\rm erg~s^{-1} cm^{-2}}
\end{equation}

with the $S_{\nu_{i}}$ the flux densities in Jy and
$\Delta\nu_{i}$ = 13.48, 5.16, 2.58, 1.0, 1.04, 1.02, 0.95, 0.65,
0.48, for $\lambda_{i}$ = 12, 25, 60, 100, 120, 150, 170, 180 and
200 ${\mu}m$, respectively\footnote{The four frequency intervals
($\Delta\nu_{i}$) relative to the IRAS wave-bands have been taken
from \citet{RR89}, while the others have been computed from Table
2.7 of \citet{Lau01}.}. We adopted a value of ${\rm H_{o} =
75~km~s^{-1}~Mpc^{-1}}$.

To show how the present sub-sample is representative of the
12$\mu$m active galaxy sample, we have computed average redshift
and completeness of the sub-sample and compared with that of the
original sample (RMS). The average redshift of the Seyfert 1's in
the present sub-sample is $\langle z \rangle =.029\pm.034$,
compared to the value of $\langle z \rangle =.030\pm.038$ of the
original sample; that of the Seyfert 2's is $\langle z \rangle
=.021\pm.018$, which is exactly the same of the value of the
original sample; that of the starburst galaxies is  $\langle z
\rangle =.034\pm.014$, compared to the value of $\langle z
\rangle =.030\pm.010$ of the original sample. The completeness
test was applied only to the Seyfert galaxies, because the
starburt nuclei of the 12$\mu$m sample are not supposed to be a
complete sample (see RMS): the present sub-sample contains 63\%
of the Seyfert 1's and 57\% of the Seyfert 2's of the original
sample down to a flux limit of 0.30 Jy at 12$\mu$m and 61\% of the
Seyfert 1's and 64\% of the Seyfert 2's of the original sample
down to 0.40 Jy. We conclude from this analysis that the present
sub-sample represents fairly well the 12$\mu$m active galaxy
sample, and there is no apparent bias.

\section{SPECTRAL ENERGY DISTRIBUTIONS}

\subsection{Average energy distributions}



Fig. 2 shows the average observed power $\nu \times F_{\nu}$ from
4400 \AA~ to 200${\mu}m$ of each class of galaxies (except for
the very few nearby spiral galaxies, for which the statistics is
poor) normalized to the 12${\mu}m$ power observed by IRAS. The
optical and near-infrared data have been corrected for aperture
sizes to represent the total fluxes (S95). The shape of the
optical-to-far-infrared spectral energy distribution (hereinafter
SED) changes from one class to another: Seyfert 1's show a general
decrease from the optical to the far-infrared with only a small
bump around the L (3.6${\mu}m$) band; Seyfert 2's show two
pronounced peaks with about the same power around the J-H bands
(1.2-1.6${\mu}m$) and at 60${\mu}m$; starburst galaxies have
again the same two peaks in their SEDs, but the far-infrared peak
is much brighter than the near-infrared peak; finally normal
galaxies show two peaks with about the same power.

To better show the far-infrared turnover, we show in Fig. 3 the
mean far-infrared SED for each class of galaxies, except nearby
spirals, normalized to the observed power at 60${\mu}m$.

The main results here are: {\it i)} Seyfert 1's have a flatter SED
shortward of 60${\mu}m$, compared to all other classes of
galaxies; {\it ii)} the two Seyfert types have virtually identical
spectra from 60 to 200${\mu}m$, while they appear different only
shortward of 60${\mu}m$ ; {\it iii)} the SED of starburst galaxies
show the steepest drop off beyond 60${\mu}m$; {\it iv)} the
normal galaxies in the sample observed show stronger emission at
wavelengths longer than 150${\mu}m$ than the other types of
galaxies.

The various types of galaxies show a sequence in the slope of the
short wavelength part  of the SED (from 12 to 60${\mu}m$) from the
very flat Seyfert 1's, through the intermediate Seyfert 2's, to
the steep starburst and normal galaxies. The flatter SED of
Seyfert 1's in the range 12-120${\mu}m$ probably arises from the
warmer dust heated by the active nucleus.

\subsection{Testing unified models of Seyfert 1's and 2's}


Unified models claim that the observational differences between
Seyfert 1's and 2's can be attributed to the different
orientation of a hypothetical dusty torus \citep[e.g.][]{anto93}.
To test this hypothesis, we have fitted in Figure~4 the average
slope of the 12-200$\mu$m far-infrared SEDs  of Seyfert 1's and
Seyfert 2's with the sum of an optically thick dusty torus seen
face and edge on respectively, from the models by \citet{gra94}
and grey-body thermal emission at 25K with inverse square
wavelength dependence of the dust emissivity (meant to represent
the extended dust emission from the galactic disk).  As shown in
Figure 4, the cool component produces about 40\% of the observed
fluxes at 60$\mu$m. Although any detailed model fitting is beyond
the scope of the present paper, there is a rough qualitative
consistency between the data and this simple model. However, {\it
in detail the model fails} because it does not predict strong
enough flux {\it differences} at 12 and 25$\mu$m as the torus
orientation shifts from face-on to edge-on. With a more realistic
{\it range of torus orientations} in Seyfert 1's and 2's, the
disagreement with observations would be even more significant.

\subsection{Spectral Decomposition of Active and Quiescent Components}

In this section we investigate if the decomposition of the
observed SEDs in physically distinct spectral components that was
proposed by \citet{RR89} (RRC) for the IRAS data can be extended
to the longer wavelength ISO results.

First, we discuss the SED as defined by the IRAS data only: if we
simply take the average flux ratios among the four IRAS
wavelengths ($F_{12{\mu}m}$: $F_{25{\mu}m}$: $F_{60{\mu}m}$:
$F_{100{\mu}m}$) of the different types of galaxies, from our
sample we obtain for Seyfert 1's: $1:2.3:10.3:21.7$, for Seyfert
2's: $1:3.3:20:29.2$, for starburst galaxies: $1:4.5:27:35$ and
for normal spiral galaxies: $1:2.4:16:30$. Subtracting the normal
galaxies ratios from those of starburst galaxies, we are able to
identify a ``warm" starburst component with ratios: $1:2:11:5$,
peaking at 60$\mu$m. As an exercise, if we add together the
ratios of Seyfert 1's with those of this warm starburst
component, we can roughly reproduce the ratios of Seyfert 2's.
This suggests that there might be a difference between the two
types of Seyfert nuclei in the star forming activity of their
host galaxies, confirming the finding that Seyfert 2 nuclei lie
preferentially in galaxies experiencing more enhanced star
forming activity compared to Seyfert 1's \citep{mai95}. It is
possible that this results from a selection effect: since Seyfert
2 nuclei are relatively weaker at 12${\mu}m$ than Seyfert 1's,
they are more likely to fall into our flux-limited sample if
their brightness at this wavelength is enhanced by strong star
formation.

We want now to extend the analysis including the longer
wavelength ISOPHOT data. Our broad-band SEDs are probably too
crude a description of the complexities of galaxies to allow a
full principal component decomposition into physically distinct
emission components. Nonetheless, we have followed RRC in using
the 12-25-60-100$\mu$m colors to identify those galaxies in our
sample which closely resemble the SEDs of the ``quiescent cirrus"
disk, the ``starburst" component, and the ``pure Seyfert"
nucleus. For each of the three types of galaxies, the normal
spirals, the starburst galaxies and the Seyfert 1's, we have
selected those objects lying in the two IRAS color-color diagrams
close (i.e., within 0.2 magnitudes) to the colors of ``pure disc",
``starburst" and ``Seyfert" components of RRC (labeled D, B and
S, respectively in Fig.1 of RRC). We have then plotted in Fig.5
the complete 12-200$\mu$m SEDs of the selected galaxies
individually and their average value for each class in Fig.6.



As shown in Fig.5, five normal spiral galaxies have nearly the
RRC colors of the ``pure cirrus disk" component: $F_{12{\mu}m}$:
$F_{25{\mu}m}$: $F_{60{\mu}m}$: $F_{100{\mu}m}$ = 1:1:12.7:40.
These are the only normal galaxies in our sample that have the
color [60 - 25]$\geq$0.8 and the color [100 - 60]$\geq$0.3.
Similarly seven starburst galaxies have IRAS colors close to the
``pure starburst" spectrum with the four IRAS fluxes scaling as
1:5:24.5:24.5. These galaxies have been selected because they are
the only starburst galaxies in our sample with 0.6$\leq$[60 -
25]$\leq$0.9 and -0.05$\leq$[100 - 60]$\leq$0.25. Seven Seyfert
1's galaxies have IRAS colors near the ``pure Seyfert nucleus" and
are the only Seyfert 1's in our sample with [60 - 25]$\leq$0.25
and [25 - 12]$\geq$0. As also confirmed from our discussion of
the two colors diagrams (see subsection 4.1), the strong
distinction between quiescent disks and starbursts remains clear
out to 200$\mu$m. The cirrus and starburst spectra probably
represent extremes of minimal and maximal recent star formation,
that tend to be found in the least and most luminous galaxies,
respectively. The pure Seyfert spectrum is rather similar to the
pure starburst spectrum between 100 and 200$\mu$m.  Both show a
relative lack of cold dust, and the Seyfert's tend to be weaker
at 120$\mu$m.

The Seyfert 2's are spread all around the IRAS color-color
diagrams.  As can be seen in Fig.5, some of them have IRAS spectra
close to the pure starburst template (we have selected four in
Fig.5). And indeed their ISOPHOT far-infrared spectra also match
the pure starburst spectrum well, since their infrared continuum
appears to be dominated from dust around star forming regions.
Those Seyfert 2's with IRAS colors like quiescent cirrus disks
also resemble pure disks in the 100--200$\mu$m region. Again it
appears that the Seyfert 2's nucleus contributes a minor fraction
of the observed far-infrared luminosity in those objects.

\section{COLOR-COLOR ANALYSIS}

\subsection{Far-infrared color-color diagrams}



We have constructed color-color diagrams with the ISO and IRAS
photometry to see how the SEDs of the different classes of
galaxies differ, and if these can be used to segregate them.
Fig.7 and 8 show, respectively, the [200 - 100] versus [60 - 25]
and the [200 - 100] versus [100 - 60] color-color
diagrams\footnote{We define color
$$ [{\lambda}_{1} - {\lambda}_{2}] = Log (F_{\nu}
({\lambda}_{1})/F_{\nu}({\lambda}_{2}))$$ where
$F_{\nu}({\lambda}_{i})$ is the flux density in Jy at
${\lambda}_{i}$. }. The averaged colors for each type of galaxies
are reported in Table 4.

From the two diagrams we conclude:
\begin{itemize}
  \item [1.] For non-Seyfert galaxies, our interpretation of the
  different positions of the different types of galaxies in Fig.8
  is that the [100 - 60] and especially the [200 - 100] colors define a correlated sequence
  running from galaxies with the weakest star formation up to the starbursts,
  in agreement with the SED behaviour as shown in Fig.3; Table 4 can be used to quantify
  this result: the [100 - 60] colors of starburst and nearby spiral
  galaxies do not overlap each other at the 1$\sigma$ level; even better, the [200 - 100]
  color is able to separate starburst from normal and normal from
  nearby spiral galaxies, while the colors of Seyfert's are
  intermediate between starburst and normal galaxies.

 \item [2.]  The star formation rate does not, however, influence the [60 - 25]
  colors; from Table 4 we note that the [60 - 25] color is
  virtually the same among starburst, normal and nearby spiral
  galaxies. The [60 - 25] color only shows a small displacement
  between Seyfert 1's and 2's, and Seyfert galaxies in general
  from normal and starburst galaxies; its average value decreases moving
  to Seyfert galaxies and becomes the lowest for Seyfert 1's
  (Table 4). However, as can be seen from the Table 4, the scatter in individual
  galaxies is comparable to the amount of this weak shift, making it marginally
  significant.

  \item [3.] Seyfert 1's and 2's are not distinguishable from
  normal galaxies, or from each other in either [100 - 60] or [200 - 100] colors;
  as Table 4 shows, for both colors Seyfert and normal galaxies
  overlap at the 1$\sigma$ level.

  \item [4.] As can be seen from both Fig. 7 and 8 and from Table 4,
  all the infrared colors of the CfA Seyfert's are more similar to
  those of normal inactive galaxies than are those of our 12$\mu$m Seyfert's.

\end{itemize}

The first result indicates that both [200 - 100] and [100 - 60]
colors decrease when star formation activity increases. We will
confirm (see subsection 4.4) that the former color also correlates
with the 60$\mu$m excess, that we consider a measure of the star
formation activity in a galaxy. The second result was already
known (e.g. S95). It comes about because recent star formation
boosts both the 25 and 60$\mu$m luminosities of galaxies in the
same proportions found in quiescent spiral galaxies. The ratio of
these two luminosities remains constant. The best explanation for
the third result is that none of the Seyfert nuclei in these
galaxies make a substantial contribution to the total fluxes
observed at wavelengths of 60$\mu$m and longward. The emission
from Seyfert nuclei is relatively stronger at 25$\mu$m, compared
with that of normal galactic disks.  This indicates that the
thermally emitting dust in Seyfert nuclei has a substantially
hotter temperature distribution than in normal galaxies, even
those with strong starbursts. Furthermore, the dust temperatures
in Seyfert 1's tend to be even higher than those in Seyfert 2's,
as was already appreciated by \citet{EM86} and \citet{EMR87}. The
fourth empirical result illustrates a general difference between
the CfA and our 12$\mu$m Seyfert samples:  in the former the
active nuclei tend to be relatively less luminous than their host
galaxies. This results in far-infrared colors in the CfA
Seyfert's which are more ``normal", i.e. similar to those of
non-active spiral galaxies. One prediction is that the
mid-infrared (10--25$\mu$m) ``compactness" [defined in
\citet{EM86} as the ratio between the flux in a small beam and
the total integrated flux] of the CfA Seyfert's is less than that
of the 12$\mu$m Seyfert's, even at a given redshift.


Figures 9 and 10 show the far-infrared colors [150 - 100] and [200
- 150] versus the IRAS colors [60 - 25] and [100 - 60],
respectively. The [200 - 150] color is able to separate starburst
from normal galaxies, as is the [200 - 100] color, while the [150
- 100] color is not. This means that the separation in mainly due

to the longer wavelength slope. Finally the separation between
CfA and non-CfA Seyfert galaxies is even more apparent in the
[200 - 150] color than in the [200 - 100] color. The CfA
Seyfert's lie in the upper part of the diagrams at [200 - 150]
$>$ -0.4, towards the position of normal galaxies, while most of
the non-CfA Seyfert's are located at [200 - 150] $<$ -0.4.

\subsection{Spectral Curvature}

Color-color diagrams can also be used to understand if the spectra
between 100--150 and 150--200 $\mu$m can or cannot be described by
simple power laws. In Fig. 11 we compare the 120 to 100$\mu$m flux
ratio with the broader wavelength baseline of 100 to 150$\mu$m.
The solid line shows the values that would be observed for pure
power laws $F_{\nu}\propto \nu^{\alpha}$ with:

\begin{equation}
\alpha = \frac{log(F_{\nu_{1}} / F_{\nu_{2}})}{log(\nu_{1}
/\nu_{2})} = \frac{log(F_{\nu_{1}} / F_{\nu_{3}})}{log(\nu_{1}
/\nu_{3})}
\end{equation}

where $\nu_{1}$, $\nu_{2}$ and $\nu_{3}$ are three generic
frequencies and  $F_{\nu_{i}}$ the correspondent flux densities.


However, most galaxy flux ratios are better fitted with curved
100--120--150$\mu$m spectra. Points lying to the right of this
line show downward spectral curvature.  Most of the sample ---
except for 4 normal galaxies and a few low-luminosity Seyfert's
--- do in fact have stronger 120$\mu$m emission than would be
predicted by a power-law interpolation between 100 and 150$\mu$m.
Relative to their 100 and 150$\mu$m fluxes, only a few galaxies
show deficits of 120$\mu$m emission. The dashed regression line
(Fig.11),

\begin{equation}
 [150 - 100] = 1.06 \times [120 - 100] - 0.09
\end{equation}

shows relatively little scatter about it. The dispersion of 0.09
dex corresponds to random errors in each flux of 16\%, not much
larger than our typical uncertainties. The line correspondent to
pure power law behaviour for these colors has a slope of 2.22
compared to the flatter slope of 1.06 fitting the data.

Similarly, the regression line fitting the [170 - 150] color
versus the [200 - 150] color is given by:

\begin{equation}
 [170 - 150] = 0.09 \times [200 - 150] - 0.03
\end{equation}

The line correspondent to a pure power law for these two latter
colors has a slope of 0.43 compared to the flatter slope of 0.09
fitting the data.

The situation appears even simpler for the 180$\mu$m flux, which
generally does fall near a power law interpolation between 150
and 200$\mu$m (as shown by the solid line in Fig. 12). Thus we
find that for most galaxies, measurements of 100, 150 and
200$\mu$m would be sufficient to predict the 120, 170 and
180$\mu$m fluxes accurately, but not with a simple power law
interpolation.


\subsection{Dust Color Temperatures}


Another important outcome that can be derived from the
color-color diagrams is the estimate of the temperature(s) that
is (are) dominating the emission in the far-infrared. To perform
this analysis we have chosen the [200 - 100] versus [60 - 25]
color-color diagram and we adopted for fitting the combination of
two grey-bodies, leaving to vary the dust emissivity law. We show
this color-color diagram separately for the different types of
galaxies in Fig. 13. The emission is broad-band: fitting it from
25 to 200$\mu$m requires a range of dust temperatures. To compare
the model with the observed colors, we have overplotted the
combination of the two grey-bodies in Fig. 13. This introduces
the minimum necessary number of fitting parameters. To fit the
long-wavelength color a steep dependence of the dust emissivity
law is needed,

\begin{equation}
 \epsilon(\lambda) \times B(\lambda,T) {\rm ~~~ with~~~~
 } \epsilon \propto \lambda^{-2}
\end{equation}

A much flatter wavelength dependence would not fit most of the
200${\mu}m$ data for any type of galaxies.

The non-Seyfert galaxies show a clear sequence of increasing
average dust temperature going from the nearest spirals up to the
most luminous starbursts. The correlation of the two colors is:

\begin{equation}
   [200 - 100] =  1.125 \times [60 - 25] - 1.252
\end{equation}

(the linear regression coefficient is 0.33 for 25 data points) or
alternately, for the [100 - 60] color:

\begin{equation}
   [200 - 100] = 1.626 \times [100 - 60] - 0.700
\end{equation}

(the linear regression coefficient is 0.73 for 26 data points)

These correlations are essentially re-statements of the
correlation first presented by S95, in their Appendix B (see
their Figure 13). The corresponding shift in grey-body
temperature that we find is the same as what they found, after
correction for the fact that their shallower wavelength
dependence of dust emissivity made their fitted temperatures
systematically higher than ours for a given observed spectrum.
Our interpretation is that higher rates of star formation raise
the average dust temperature, shifting the peak of the thermal
dust emission to higher frequencies, and tilting the entire
far-infrared spectrum toward the blue. Splitting the
100--200$\mu$m spectrum into two pieces, we find that most of
this curvature occurs between 150 and 200$\mu$m.

In the normal spirals belonging to the 12${\mu}m$ galaxy sample
and observed by \citet{S99} and by \citet{A98}, dust at even lower
temperature (15 K) is detected, confirming the previous finding
of these latter authors.

Most of the Seyfert 1's far-infrared data occupy a region that can
be fitted by the mixture of two grey-bodies with ${\rm
T_{min}=22K}$ and ${\rm T_{max}=55-70K}$. The three nearby CfA
Seyfert 1's galaxies (NGC~3227, NGC~5033, NGC~4051, with data
taken from PGRE) with an average redshift of z=0.0031, show the
highest value of the [200 - 100] color, requiring a lower minimum
dust temperature (the fit in Fig. 13a gives ${\rm T_{min}=18K}$).

Almost all Seyfert 2's far-infrared data (Fig. 13b) can be fitted
by the mixture of two grey-bodies with ${\rm T_{min}=25K}$ and
${\rm T_{max}=55-65 K}$. They show a somewhat narrower
temperature distribution compared to Seyfert 1's, and do not
extend to the higher values. Again the four reddest CfA Seyfert
2's --- NGC~4388, NGC~3982, NGC~1143/44 and NGC~3079 --- (PGRE)
with the coolest dust (${\rm T_{min}=20K}$), have an average
redshift of 0.011, as opposed to 0.019 for all the other Seyfert
2's plotted.

Comparing the three classes of galaxies, there is a trend with
the most active objects (Seyfert 1's) have the widest temperature
range (22--70K); Seyfert 2's are intermediate, while starbursts
have the warmest low temperature component (30K), together with
the coldest high temperature one (60K).

A similar result has been found by \citet{K97} analyzing the
ISOPHOT data of the three interacting galaxies Arp~244, NGC~6240
and Arp~220: the increase of the highest dust temperature
component is accompanied by the decrease of the lowest dust
temperature component they detect. In four narrow-line Seyfert 1's
\citet{PC99} found that the lowest dust temperature is in the
range 20-40K.

We are aware that there could be an additional selection effect in
nearby galaxies if there is very extended emission from the galaxy
discs. Measurements not performed in mapping mode could have lost
some of the long wavelength emission. The detection of very cold
dust only in nearby objects could thus be explained with the very
low surface brightness of its thermal emission. This dust
component has been found by \citet{A98} to be surprisingly
extended. We cannot exclude therefore that also for our objects
there might be emission from cold dust (T $\sim$ 15K) that
escaped detection with our observations.

\subsection{The far-infrared turnover versus the 60$\mu$m excess}

In this subsection we want to relate the strength of the star
formation activity in a galaxy with the steepness of the
far-infrared turnover, measured by the [200 - 100] color, with
the aim to see how the different classes of galaxies behave and
if there is a separation between galaxies with and without active
nuclei.

We have chosen as the indicator of enhanced recent star
formation, which warms dust around HII regions, the 60$\mu$m
``excess" as the ratio of the observed 60$\mu$m flux to the flux
that a source would have at 60$\mu$m from power-law interpolation
of the flux between 12 and 100$\mu$m. We have already seen (see
subsection 4.1) that the starburst galaxies of our sample are
characterized by a low value of their [200 - 100] color. In
Fig.14 we plot the [200 - 100] color index as a function of the
60$\mu$m excess. We find a correlation between the steepness of
the far-infrared (100--200$\mu$m) turnover and the strength of
the 60$\mu$m ``excess".  While this diagram does not perfectly
separate the galaxies of different classes, it nevertheless shows
that they cluster preferentially in different regions of the
diagram. Seyfert 1's (excluding six objects of the CfA sample)
cluster in a no-60$\mu$m excess region with a color [200 - 100]
$<$ 0. The starburst galaxies cluster in the central area of the
diagram and have all 60$\mu$m excess. Normal galaxies and nearby
spirals have no 60$\mu$m excess (except 2 objects) and a color
$-0.5<[200 - 100]<+0.5$. Seyfert 2's are widely spread all over
the diagram, but with a 60$\mu$m excess generally higher than
Seyfert 1's. Four especially infrared-luminous galaxies ---
MK~273, Arp~220, MK~938 and FSC05189-2524 lie in the ``starburst"
zone. In fact detailed mid-infrared spectroscopy (in the range
3-30$\mu$m) suggests that much of the total infrared luminosity
in these extreme objects is indeed powered by star formation
\citep{Gen98}. As we already noted, many of the CfA Seyfert
galaxies have higher excess and/or higher [200 - 100] color, with
respect to the other Seyfert galaxies. This is presumably because
their faint Seyfert nuclei are too feeble to influence the
overall far-infrared continuum emitted by their bright host
galactic disks.


As stated above, the 60$\mu$m excess is measured with respect to
the underlying 12 and 100$\mu$m fluxes, which are  dominated by
the ``cirrus" component  powered by the integrated stellar
luminosity in the galaxy. When the 60$\mu$m bump is especially
strong, the warm dust associated with recent star formation even
contributes to the observed 100$\mu$m flux, making a relative
excess at that wavelength compared with the emission from cold
dust at longer wavelengths. The result is that the more actively
star-forming galaxies, such as infrared-luminous starbursts, have
the sharpest far-infrared turnover. They show the {\it steepest}
drop off from 100 to 200$\mu$m, with an average slope over that
wavelength range of $\alpha_{100-200{\mu}m}$ = + 2.2, compared to
$\alpha_{100-200{\mu}m}$ = + 1.4, $\alpha_{100-200{\mu}m}$ = +
1.6 and $\alpha_{100-200{\mu}m}$ = + 0.8 for Seyfert 1's, 2's and
normal galaxies, respectively (see Table 4).

A simple least squares fit to the data show a correlation of the
color [200 - 100] with the 60$\mu$m excess with a linear
regression coefficient of R=-0.48 (for 83 objects), corresponding
to a probability of 99.9996\%. This correlation improves by
excluding the Seyfert 1's (R=-0.61 for 65 objects).

We suggest that that the diagram shown in Fig.14 can be used to
separate the {\it starburst dominated} objects from the {\it AGN
dominated} ones. Objects located in the upper right part of the
diagram are ``more" starburst dominated, while those at the left,
having a fainter excess, are the AGN dominated objects. We
suggest that starburst activity in galaxies, i.e. with high rates
of current star formation, results in excess emission in the
60$\mu$m band accompanied by a general heating of the galactic
ISM and thus a decrease of the [200 - 100] color.

We note that many of the Seyfert galaxies of the CfA sample are
shifted upwards in Figure 14 with respect to the bulk of the
12$\mu$m Seyfert's. Again we interpret this as an indication that
their stronger long-wavelength emission arises from the dominance
of quiescent disk dust.

\section{LUMINOSITY ANALYSIS}
\subsection{Luminosity Correlations}


In this subsection we analyze the correlations of the 200$\mu$m
luminosity with the luminosity in each of the IRAS bands and the
total mid- to far-infrared luminosity. Although diagrams which
correlate one luminosity with another in the same object should
be used with caution, we have a large sample with good selection
criteria in which the luminosities span over 4 orders of
magnitude.  In Figures 15 and 16 we show such correlations and
give the formulae for the best-fitting regression lines. As
expected, the slopes in the correlations of the 200$\mu$m
luminosity against the 25 and 60$\mu$m luminosities are flatter
than those against the 12 and 100$\mu$m luminosities (see also
S95). This is because the 25 and 60$\mu$m luminosities are
preferentially elevated in those galaxies with more active recent
star formation.  The 100$\mu$m and 200$\mu$m emission, in
contrast, are relatively stronger in the quiescent disk emission
(see below), and correlate almost linearly with each other.

These figures are a logical extension of the findings of S95.
They showed that galaxies with higher bolometric luminosity are
relatively brighter in the 25 and 60$\mu$m bands, due to the
elevated ``starburst" component in them, while they appear
relatively fainter at both shorter (12$\mu$m) and longer
(100$\mu$m) wavelengths. The ``pivot points" at which the relative
luminosity neither increases nor decreases, occur around 12 and
100$\mu$m. This is our interpretation of why the 12 and 100$\mu$m
luminosities have a correlation with the bolometric luminosity
with a flatter slope (1.06 and 1.13, respectively) compared to the
slope of the correlations of the 25 and 60$\mu$m luminosities
with the bolometric luminosity ($\sim$ 1.2)(see Fig.7 of S95). For
wavelengths much shorter than 12$\mu$m or longer than 100$\mu$m
we therefore expect the luminosity to track less than linearly
with the bolometric luminosity. That is what these figures are
showing.

It is not particularly surprising that the 200$\mu$m luminosity
correlates most closely with the 100$\mu$m luminosity, given that
these are the two closest wavelengths. The fact that the Seyfert
galaxies (of both types 1 and 2) lie on the same track as the
non-Seyfert's once again indicates that the Seyfert nucleus makes
a negligible contribution at these wavelengths. Since the
200$\mu$m luminosity is dominated by the ``quiescent cirrus"
component, it does not increase as fast as the 25 or 60$\mu$m
luminosities, which are strongly effected by starbursts. Thus the
slopes of the correlation between the 200$\mu$m luminosity and
those at these latter wavelengths are much flatter than 1 (0.83
and 0.86 respectively).

The 200$\mu$m luminosity even decreases less rapidly than the
total luminosity.  This is seen by the flat slope of only 0.79 in
Figure 16, where the X-axis is the ``total" mid- to far-infrared
luminosity, defined in equation (1) and given in Table 3. We note
that a similar slope (0.87) is relating the optical blue
luminosity with the bolometric luminosity (see Fig.10 of SM95),
confirming the above interpretation.


\subsection{Color-luminosity diagrams}



As discussed above, it is believed that the more luminous
non-Seyfert galaxies have higher light/mass ratios.  They are
luminous not so much because they are bigger or more massive, but
principally because they have higher proportions of recently
formed stars, which are highly luminous. A higher proportion of
their total luminosity emerges in the far-infrared, and in
particular from the warmed dust grains which emit strongly at
25--60$\mu$m-- what we describe as the ``starburst" component
(e.g. S95). One consequence of this is that the most actively
star-forming galaxies should have the sharpest far-infrared
turnovers, as shown in subsection 4.4, because their dust has a
warmer temperature distribution, as seen is subsection 4.3. Thus
we expect a systematic trend of infrared SEDs with total
luminosity. \citet{S95} already showed that the higher the
bolometric luminosity of a galaxy, the higher its ratio of
starburst/cool dust components.

Indeed we do find that the far-infrared turnover becomes
systematically sharper in the more luminous galaxies. This can be
seen in the inverse correlation between the color index [200 -
100] and the 12$\mu$m luminosity, shown in Fig. 17. In this
figure we see that this relation holds in general, irrespective
of the type of the galaxy, with a regression coefficient of
R=-0.48 for 81 objects, for all the observed galaxies in the
12$\mu$m sample, excluding the 3C and PG quasars. The fit to the
non-Seyfert galaxies only is:

\begin{equation}
  [200 - 100] = 13.513 - 0.317 \times Log (L_{12{\mu}m})
\end{equation}

(the linear regression coefficient is -0.52 for 26 data points).

We also plot in Fig.18 the inverse correlation between the color
index [200 - 100] and the 60$\mu$m luminosity, which gives a
regression coefficient of R=-0.48 for 83 objects, for all the
galaxies in our sample, excluding the 3C and PG quasars. The fit
for the non-Seyfert's only gives:

\begin{equation}
 [200 - 100] = 14.983 - 0.346 \times Log (L_{60{\mu}m})
\end{equation}

(the linear regression coefficient is -0.63 for 26 data points)

The four most noticeable outliers are the two 3C objects and the
two PG quasars, which have unexpectedly red [200 - 100] colors .
There is therefore an indication that even at low redshift (z
$\leq$ 0.2) quasars (either the optically selected PG and the
radio selected 3C) might have an excess in cold dust emission.

In Figures 19 and 20, we separate this color-luminosity trend into
the slopes between 100 and 150$\mu$m and 150 to 200$\mu$m.
Clearly the trend is due entirely to the steepening of the
200--150$\mu$m slope in the more luminous galaxies, which is the
same whether the luminosity is measured at 12 or at 60$\mu$m. The
100--150$\mu$m slope in fact shows no significant luminosity
dependence.



\section{SUMMARY}

New ISOPHOT photometry of a large sample of nearby active and
normal galaxies shows that:
\begin{itemize}
\item[i)] The 60-200${\mu}m$ SEDs of active and normal galaxies
are similar.  The 100 to 150 to 200$\mu$m spectra have
significant curvature, generally downward.  The trends are
sufficiently systematic that measurements at these wavelengths
can give accurate predictions of the fluxes at the intermediate
wavelengths of 120, 170 and 180 $\mu$m.
\item[ii)] In the non-Seyfert galaxies, higher luminosities (presumably
generated by higher rates of recent star formation) are correlated
with hotter average dust temperatures at all wavelengths from 12
to 200$\mu$m.  In particular, the 150 to 200$\mu$m slope steepens
in more luminous galaxies.
\item[iii)] The decomposition of the observed SEDs into physically
distinct spectral components has been extended from the IRAS data
(RRC) to our ISO results: a strong distinction is apparent out to
200${\mu}m$ between the the quiescent disk component and the
starburst component. We suggest that these components represent
the extremes of minimal and maximal recent star formation, found
in the least and most luminous galaxies, respectively.
\item[iv)] The [200 - 100] versus [100 - 60] color-color diagram
is able to separate the three types of galaxies: starburst
galaxies, normal galaxies and nearby spirals, while Seyfert's are
at intermediate locations.  The large-aperture 60--200 $\mu$m
spectra of Seyfert galaxies we measured are dominated by the
emission from the host galaxy.  They therefore depend on the
relative contributions of ``starburst" and ``quiescent disk"
emission in the galaxy, and are independent of the presence of an
active nucleus.
\item[v)] The [200 - 100] vs [60 - 25] color-color
diagram shows that a mixture of two black-bodies with a warm and
a cold component is able to fit all the far-infrared data, only
if we assume the dust emissivity law proportional to the inverse
square of the wavelength. Comparing active galaxies, Seyfert 1's
show the larger range in temperature (22-70K), Seyfert 2's are
intermediate, while starburst galaxies have the narrower range in
dust temperature (30-60K). Nearby normal spirals and some of the
CfA Seyfert galaxies show very cold dust components (15-18K).
\item[vi)] A correlation is found between the slope of the far-infrared
turnover (as measured from the [200 - 100] color) and both the
60$\mu$m excess and the 12 and 60$\mu$m luminosities. We have
confirmed and extended the finding of S95 that increasing star
formation rates cause the more luminous non-Seyfert galaxies to
have relatively stronger 25--60$\mu$m emission, resulting in
systematically bluer far-infrared colors. We suggest that the
plane defined by the [200 - 100] color versus the 60$\mu$m excess
can be used to separate {\it starburst dominated} galaxies from
{\it AGN dominated} ones.

\end{itemize}

\acknowledgments This research was funded in Italy from the
Italian Space Agency (ASI). This work benefited from extensive
use of the NASA Extragalactic Database (NED) and from the help
given by the IPAC staff in planning the ISO observations. We
thank Brian Rush and Nao Suzuki for help in preparing the
observing command templates. We also thank the anonymous referee,
whose thorough comments helped in improving this article.



\appendix
\section{DATA OBTAINED WITH THE ISOPHOT C100 ARRAY CAMERA}

Some C100 observations, namely for objects FSC00521-7054,
NGC~7674, MCG+1-33-36, NGC~262, FSC03362-1642, ESO253-G3,
NGC~1365, NGC~4501, NGC~4922A/B were performed in staring mode,
however only for the latter three an adjacent empty sky position
was observed to measure the sky background. In the former cases
sky background was estimated as follows: we assume that the
sources are point-like and therefore only the central pixel (\#
5) contains the source flux. The adjacent pixels were averaged to
get a background value, which was then subtracted from the signal
of the central pixel. This procedure is, however, affected by
large uncertainties since it requires a good flat-fielding of the
array because of its unstable behavior and in two cases, for
NGC~7674 and NGC~262, the values inferred are much lower than the
corresponding IRAS photometry. Note that the uncertainties quoted
here are only the statistical ones. The results of the C100
observations are reported in Table 5.




\clearpage


\begin{figure}
\plotone{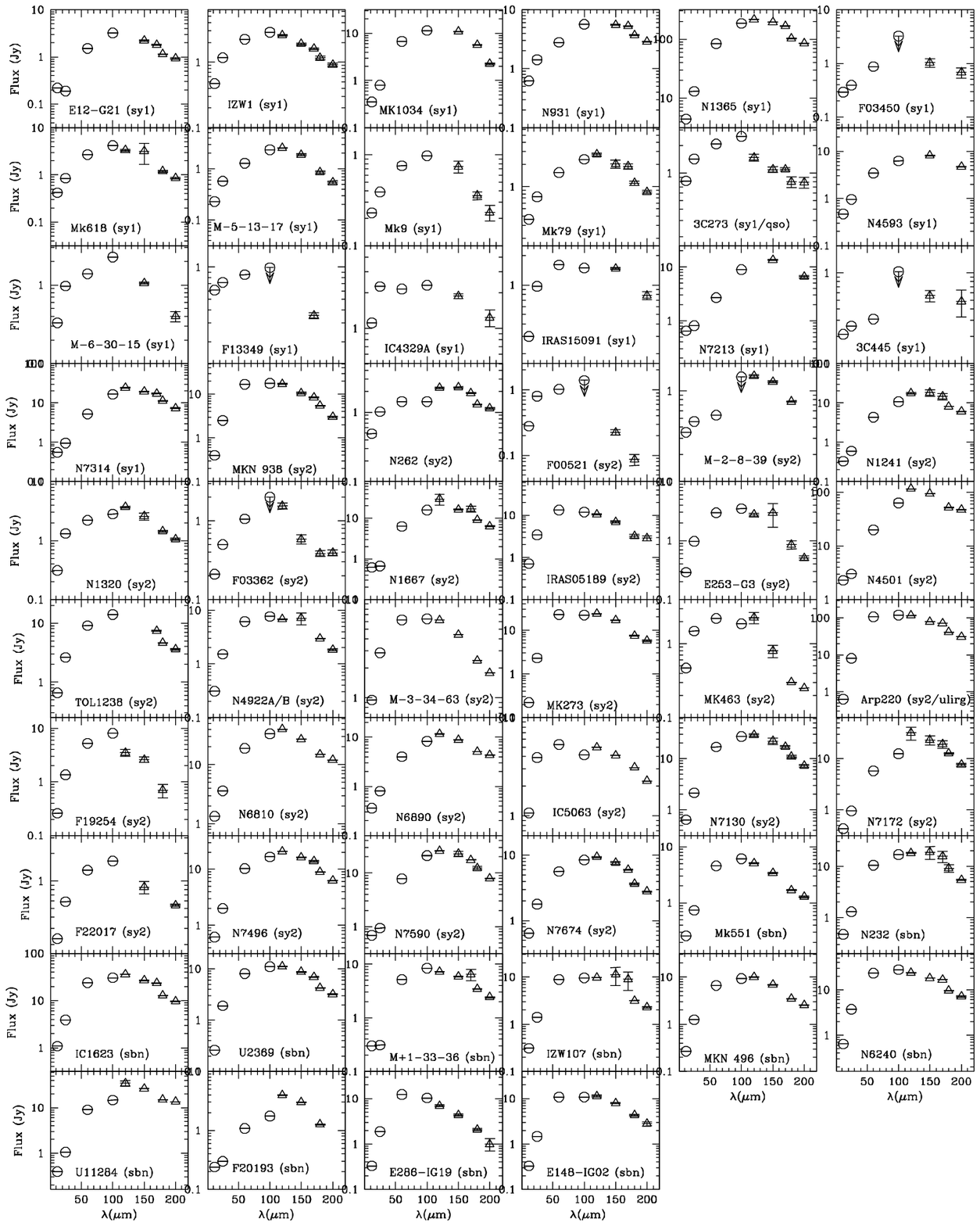} \caption{ISOPHOT data (open triangles) and IRAS
fluxes (open circles) from RMS. \label{fig1}}
\end{figure}
\clearpage

\begin{figure}
\plotone{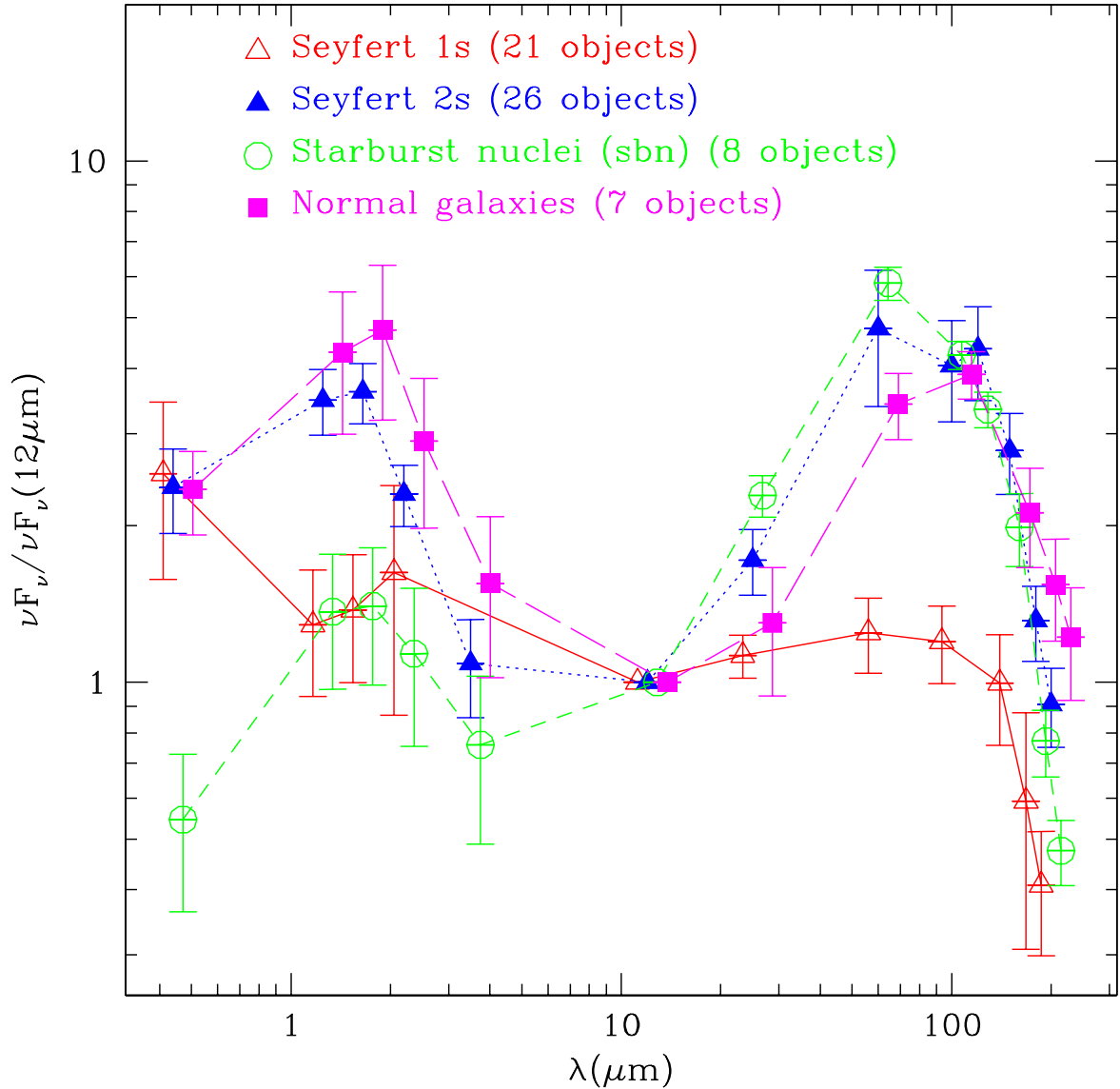} \caption{Combined SEDs of active and normal
galaxies of the 12$\mu$m sample from 4400\AA~ to 200$\mu$m,
normalized to 12${\mu}m$ (Note that the number of objects plotted
here is smaller than the total number of galaxies because some of
then do not have near-infrared and/or optical data in S95. Note
also that the 170$\mu$m band has been excluded from the average
values in this figure and in the following Fig.3,4 and 6 because
the statistics is poor, as can be seen from Table 2).
\label{fig2}}
\end{figure}
\clearpage

\begin{figure}
\plotone{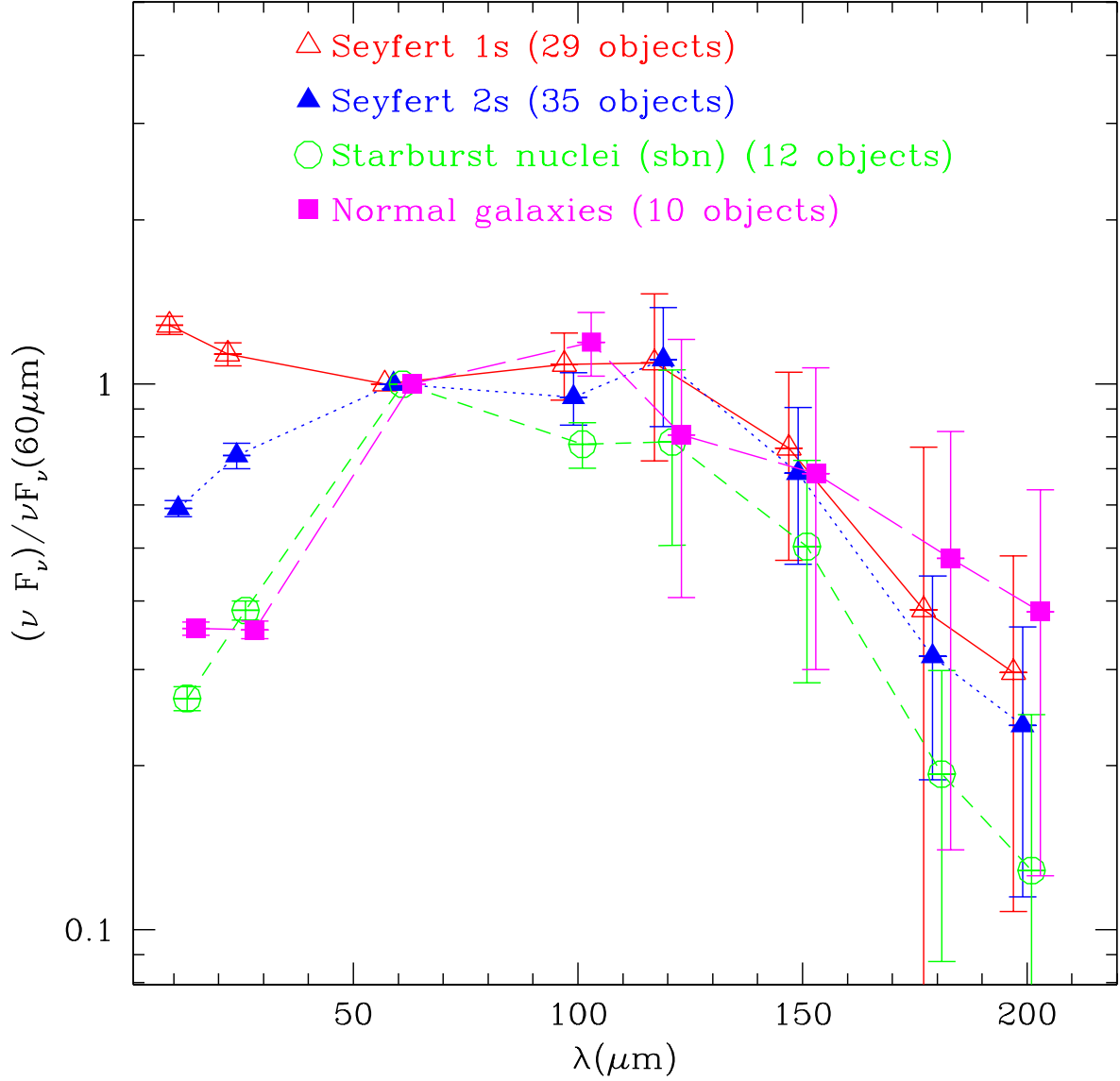} \caption{The average SED normalized to
60${\mu}m$ of galaxies belonging to the 12${\mu}m$ galaxy sample.
The nearby spirals class has been excluded, because of poor
statistics. \label{fig3}}
\end{figure}

\clearpage

\begin{figure}
\plotone{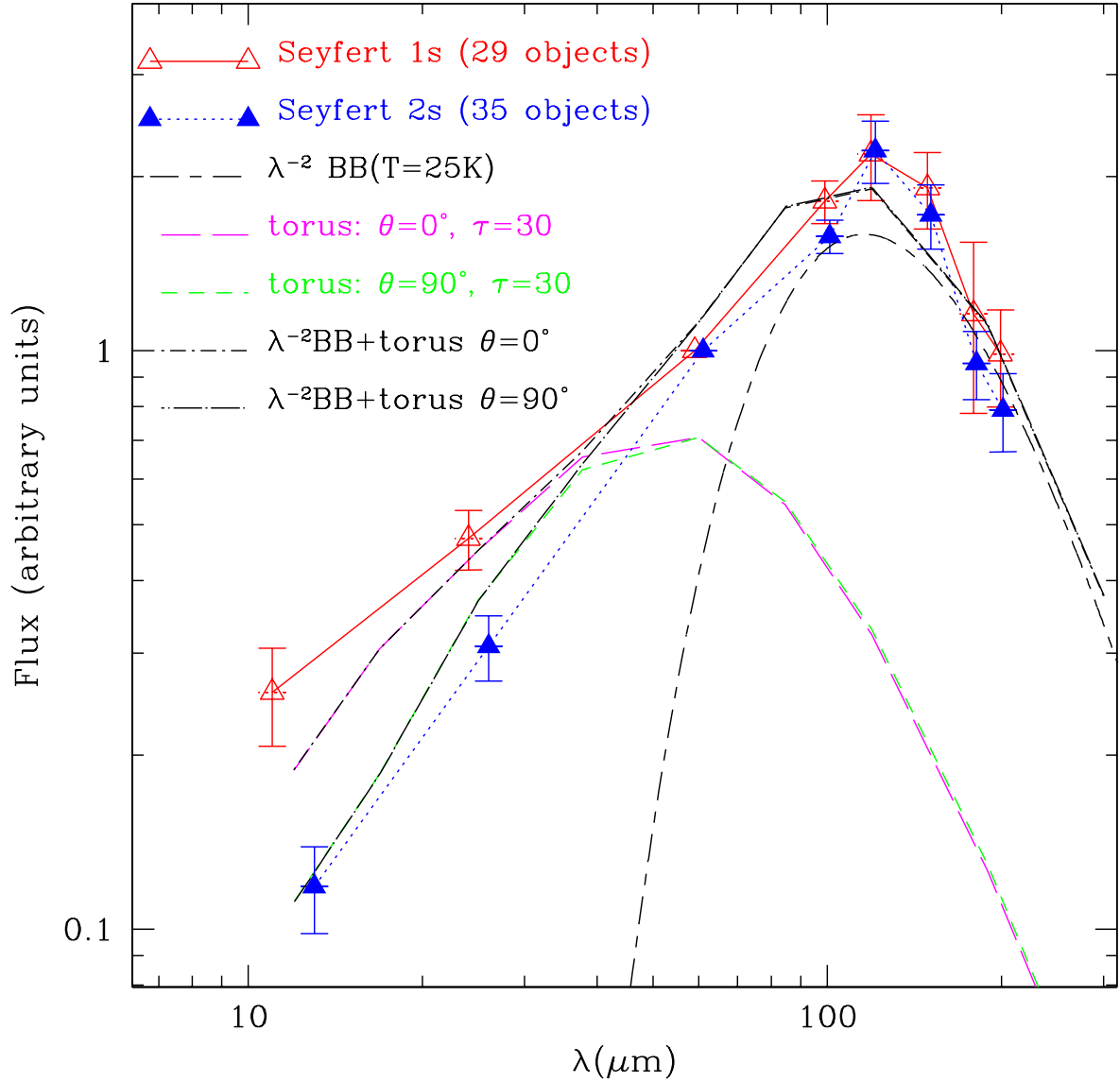} \caption{Comparison between the average Seyfert
1's (open triangles) and 2's (filled triangles) SEDs. The short
dashed line shows the edge-on model ($\theta=90\degr$,
$\tau=30$); the long dashed line, the face-on model
$\theta=0\degr$, $\tau=30$) from \citet{gra94}. The short and
long dashed line shows the component of host-galaxy disk emission
which was added to the torus models to match the overall SEDs of
the Seyfert galaxies--it is approximated as grey-body emission at
T=25K. \label{fig4}}
\end{figure}
\clearpage

\begin{figure}
\plotone{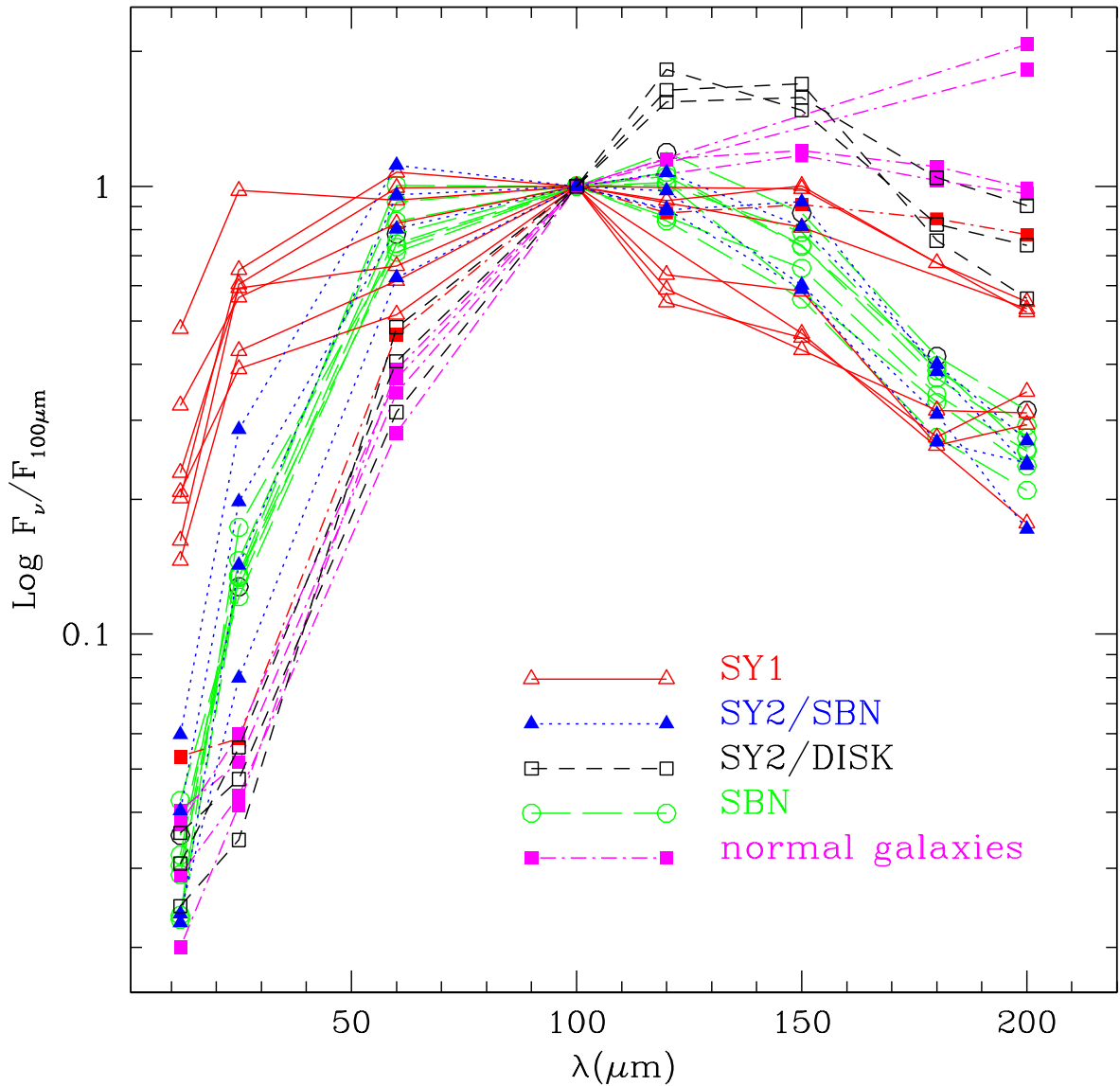} \caption{The flux densities normalized to the
100 $\mu$m flux density of the galaxies chosen to represent the
different emission components. The chosen normal spiral galaxies
are: NGC~7624, NGC~7083, UGC~2936, NGC~134 and NGC~5194 (M~51);
the starburts galaxies are: IC~1623, MK~496, IZw~107, MK~551,
UGC~2369, ESO~148-IG02 and NGC~6240; the Seyfert 1's are:
FSC15091-2107, 3C~273, IC~4329A, MCG-6-30-15, NGC~4151, NGC~5548
and MK~817. Finally the Seyfert 2's with starburst SED are:
FSC~05189-2524, MK~938, NGC~7130  and NGC~4922; while those with
disk SED are: NGC~1241, NGC~4501 and NGC~3079. Note that the data
at 170$\mu$m have been omitted because available only for few
objects. \label{fig5}}
\end{figure}
\clearpage

\begin{figure}
\plotone{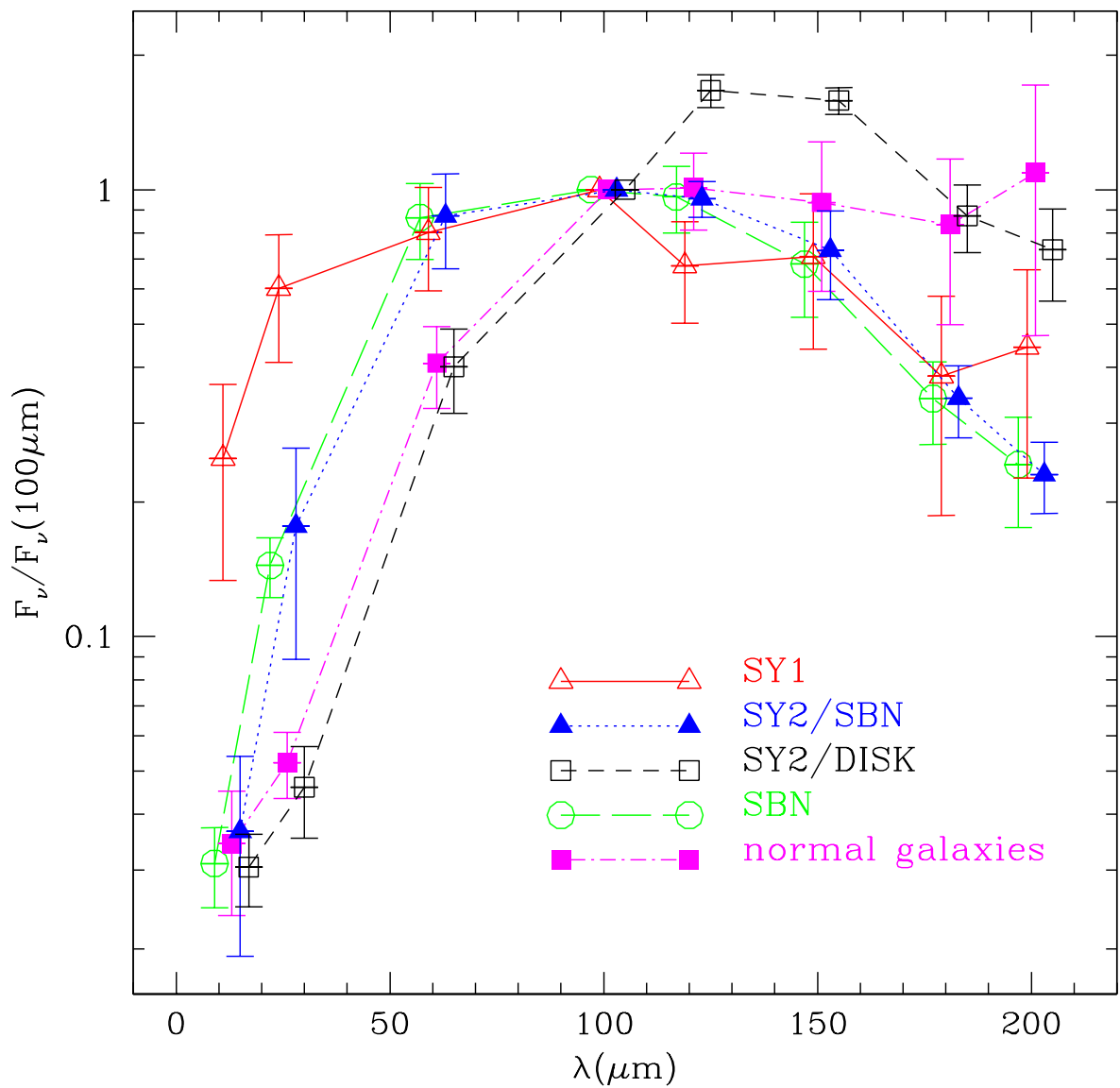} \caption{The average flux densities for each
type of galaxies representing different emission components
(``pure" Seyfert, starburst, disk emission, starburst-like
Seyfert 2 and disk-like Seyfert 2). These are the averages of the
data plotted in Figure 5, again normalized to the 100$\mu$m flux
density.\label{fig6} }
\end{figure}
\clearpage

\begin{figure}
\plotone{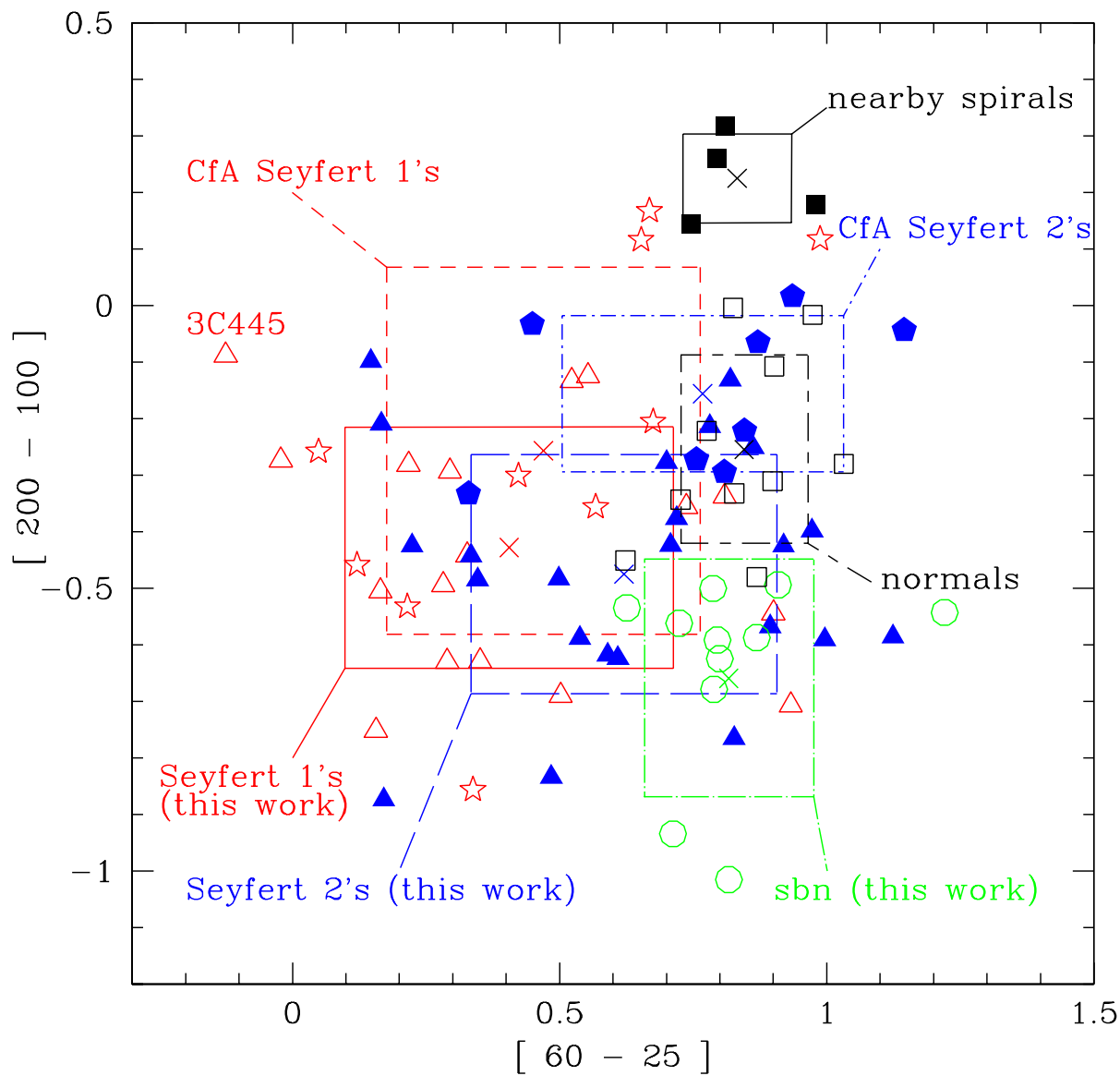} \caption{The [200 - 100] versus [60 - 25]
color-color diagram of the various galaxies belonging to the
12${\mu}m$ galaxy sample. Open triangles represent Seyfert 1's,
open stars are CfA Seyfert 1's (PGRE), filled triangles are
Seyfert 2's, filled pentagons are CfA Seyfert 2's (PGRE), open
circles are starburst nuclei, open squares are normal galaxies
and filled squares are nearby spiral galaxies. The mean colors of
each class of galaxies are indicated by crosses, with their
1$\sigma$ dispersion shown by boxes. \label{fig7}}
\end{figure}
\clearpage

\begin{figure}
\plotone{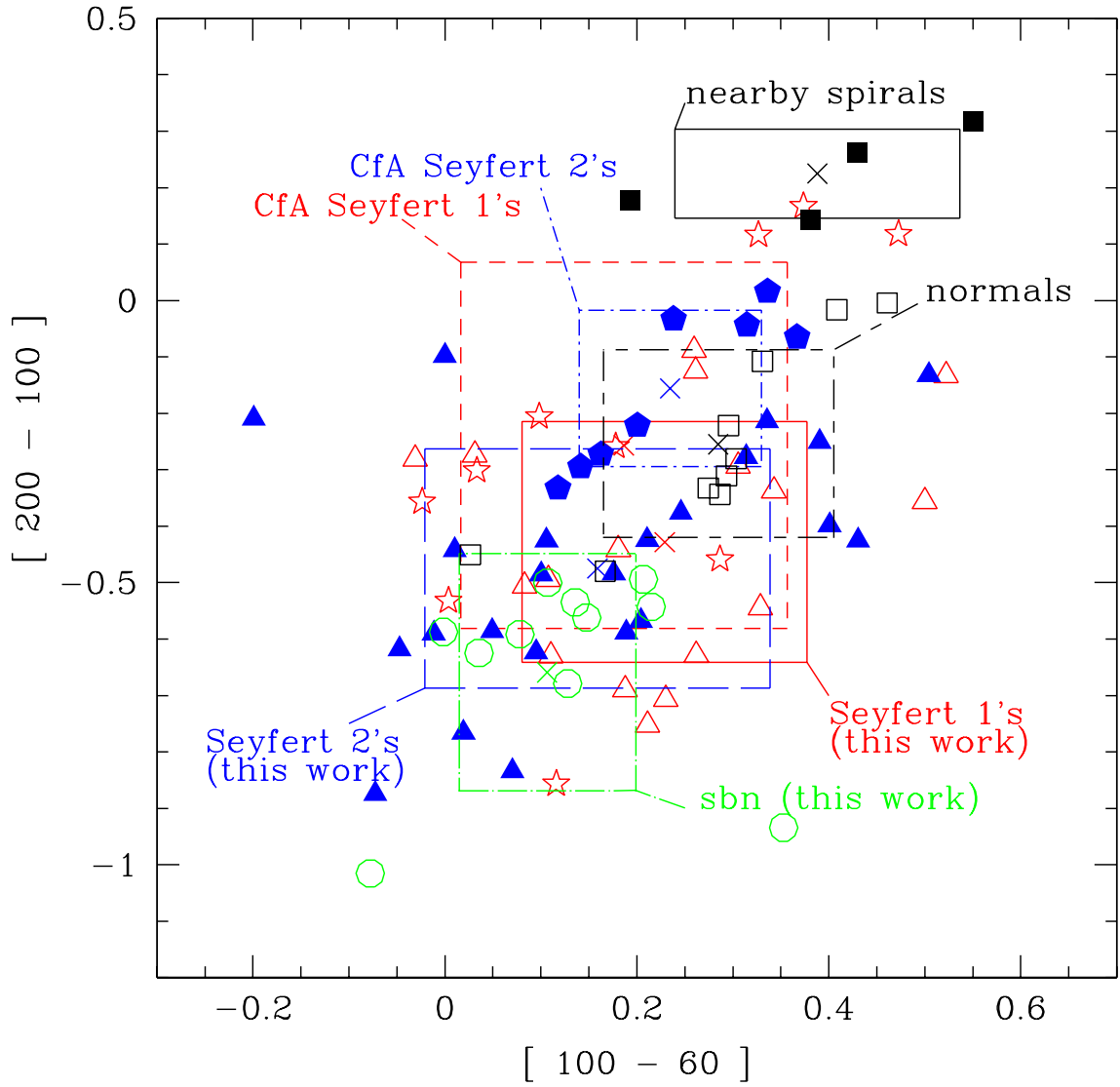} \caption{The [200 - 100] versus [100 - 60]
color-color diagram of the various galaxies belonging to the
12${\mu}m$ galaxy sample. We refer to Fig.7 for the notation.
\label{fig8}}
\end{figure}
\clearpage

\begin{figure}
\plotone{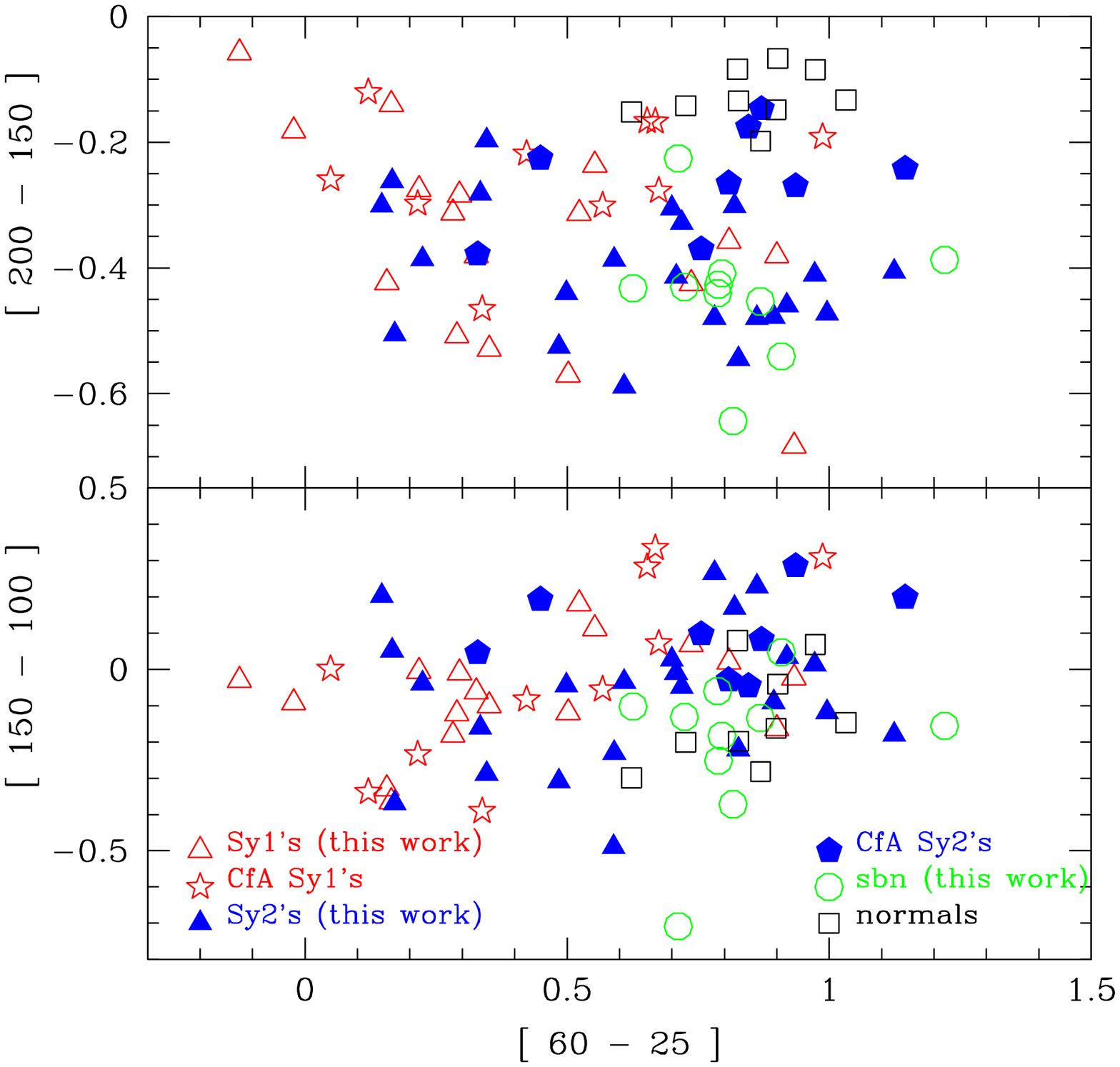} \caption{The far-infrared colors [150 - 100]
(lower panel) and [200 - 150] (upper panel) versus the IRAS color
[60 - 25]. \label{fig9} }
\end{figure}
\clearpage

\begin{figure}
\plotone{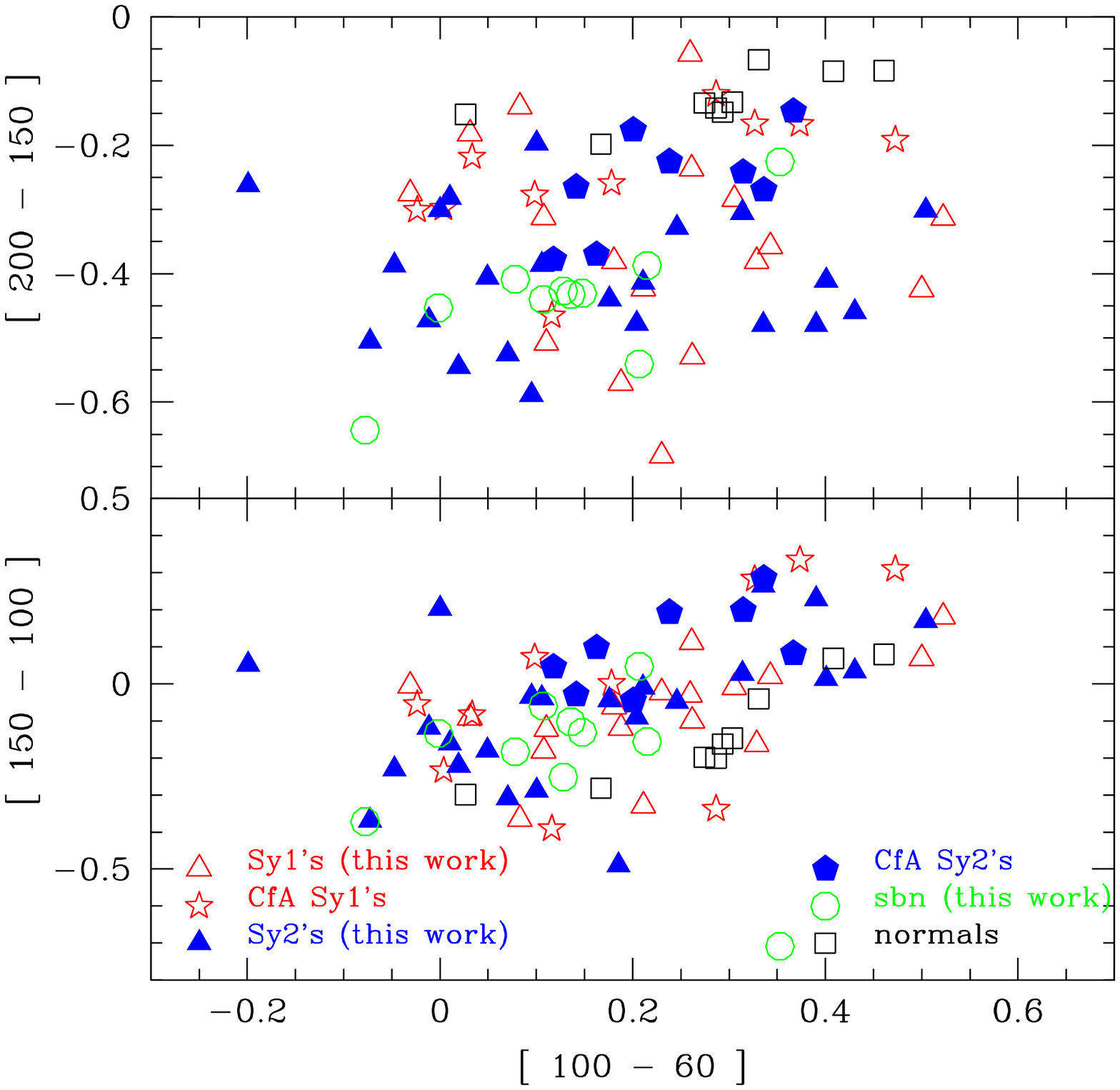} \caption{The far-infrared colors [150 -
100](lower panel) and [200 - 150] (upper panel) versus the IRAS
color [100 - 60] \label{fig10}}
\end{figure}
\clearpage

\begin{figure}
\plotone{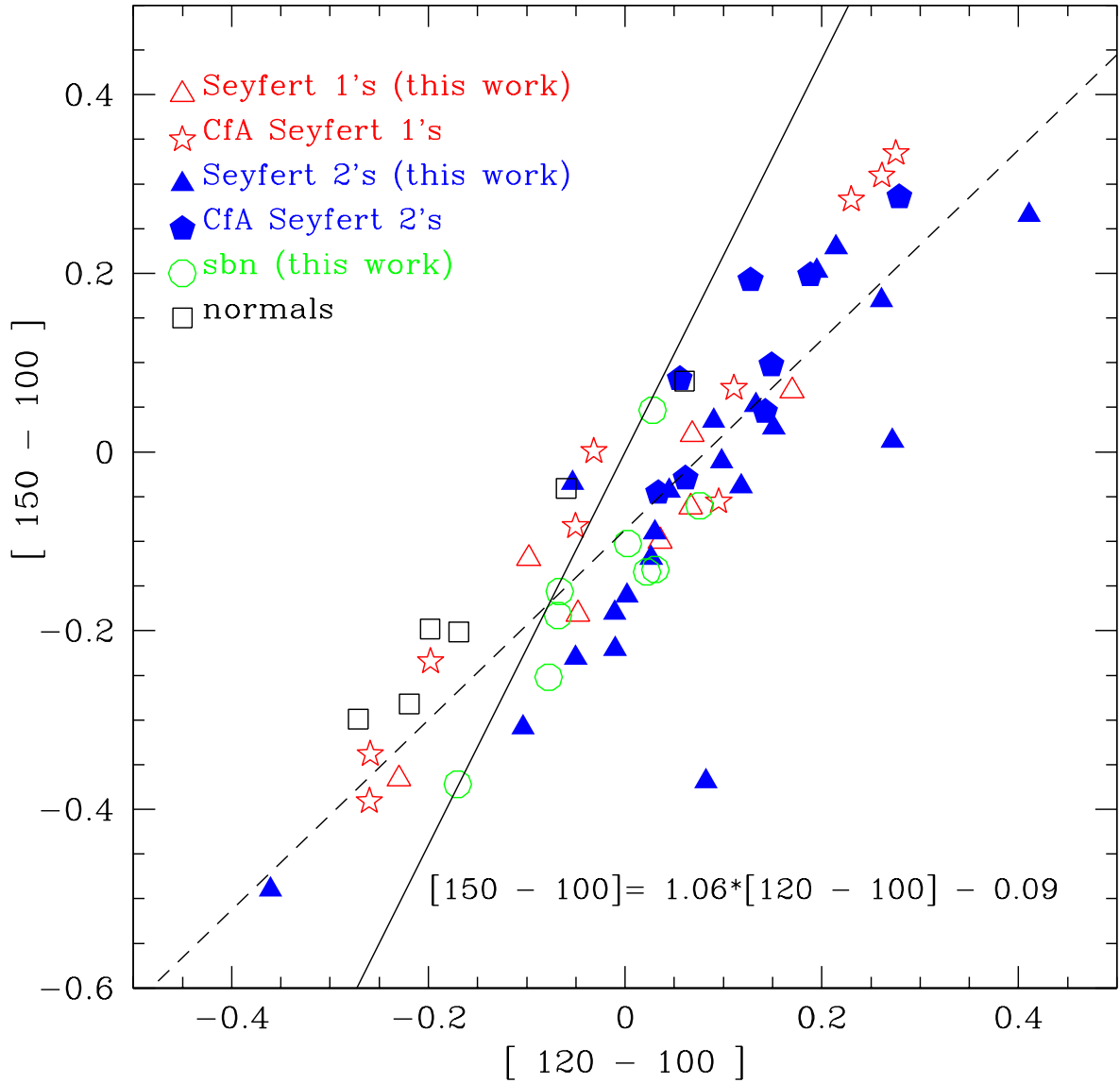} \caption{The [150 - 100] versus the [120 -
100] color-color diagram. The dashed line represents the least
squares fit to the data, the solid line gives the locus of a pure
power law spectral dependence of flux densities. \label{fig11}}
\end{figure}
\clearpage

\begin{figure}
\plotone{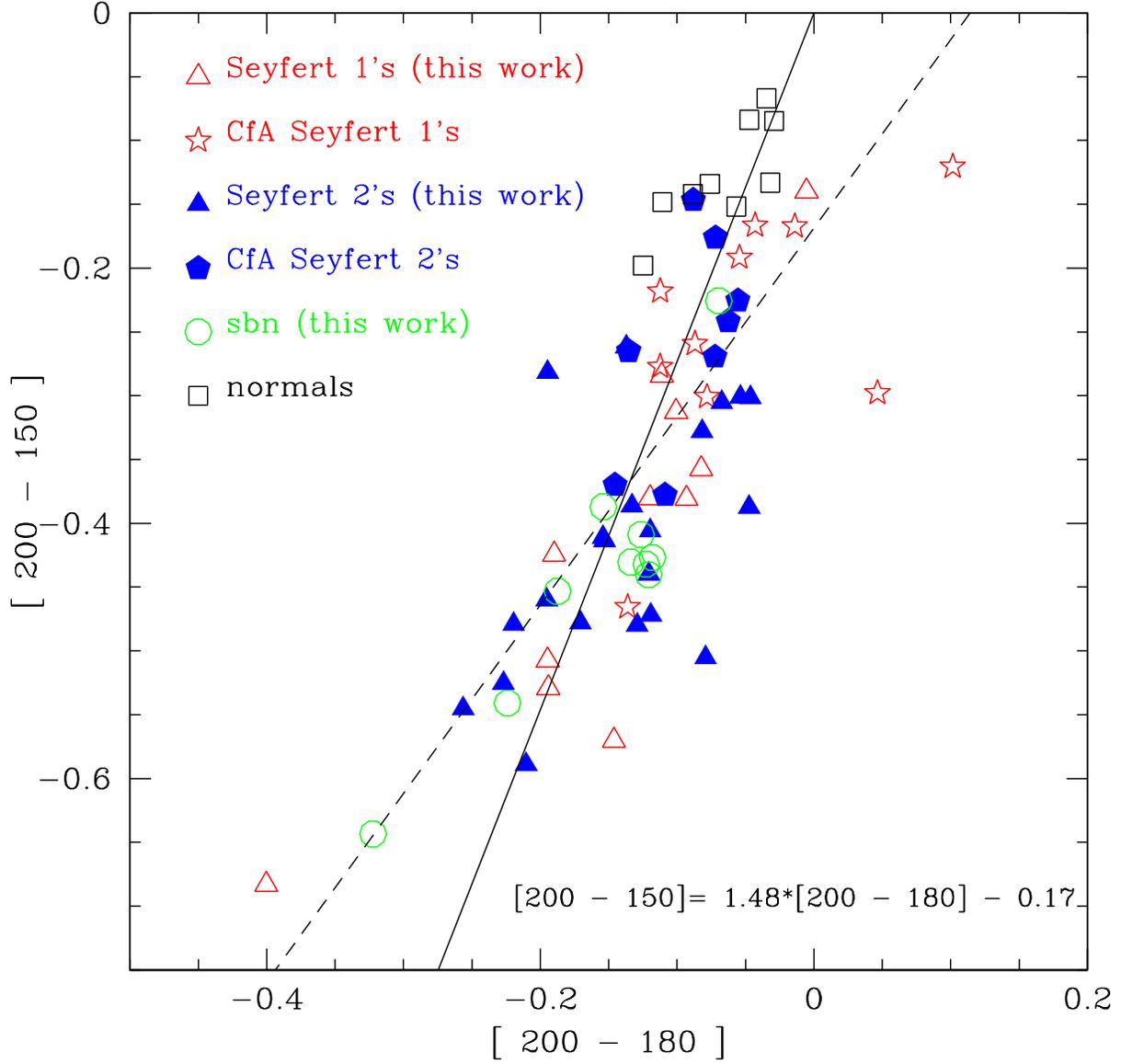} \caption{The [200 - 150] versus the [200 -180]
color-color diagram. Dashed and solid line as for the previous
figure. \label{fig12}}
\end{figure}
\clearpage

\begin{figure}
\plotone{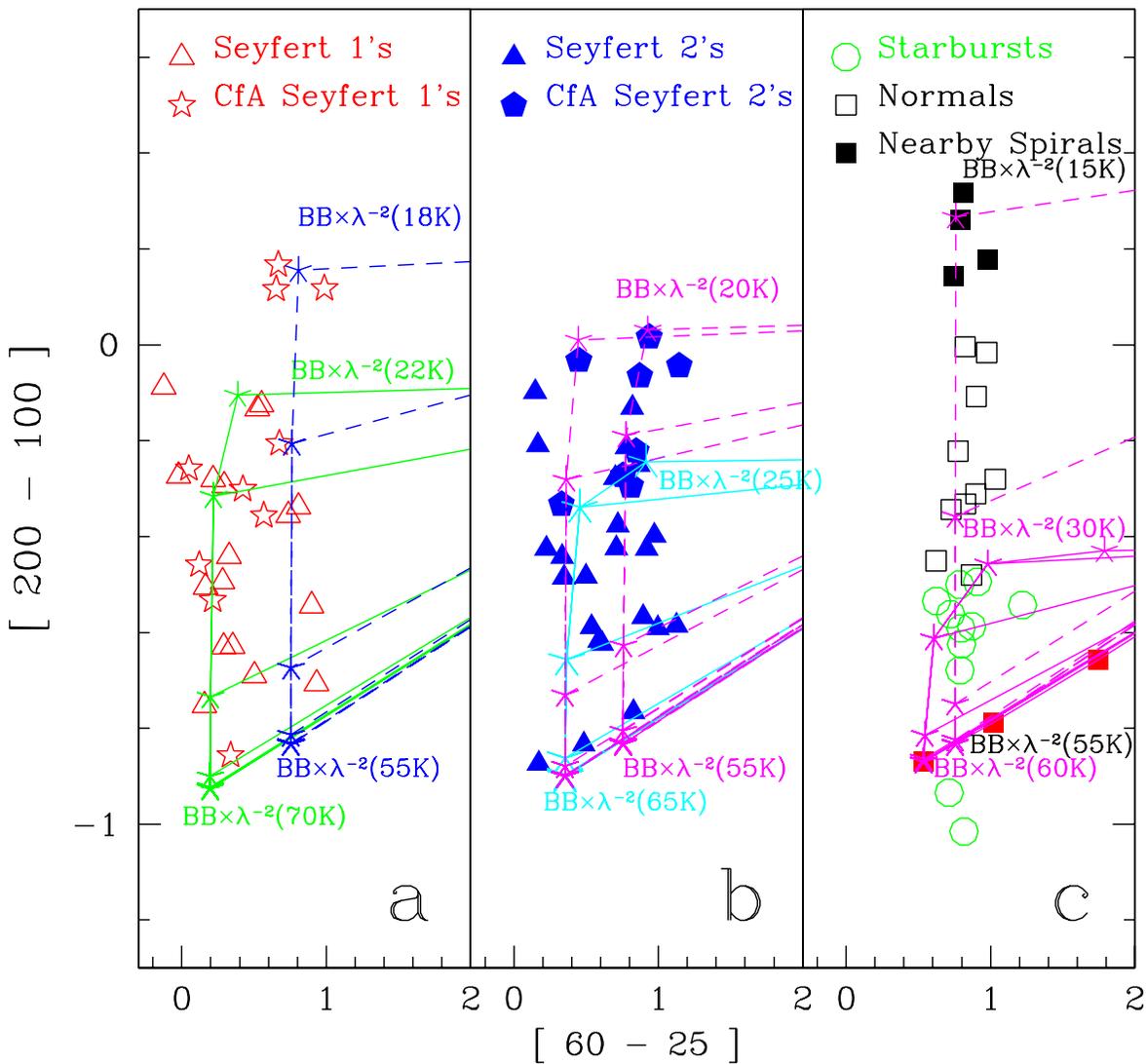} \caption{The [200 - 100] versus [60 - 25]
color-color diagram of the various classes of galaxies belonging
to the 12${\mu}m$ galaxy sample. a: Seyfert 1's galaxies. b:
Seyfert 2's galaxies. c: starburst and normal galaxies. The lines
represent the models made of the mixture of two black-bodies at
the quoted temperatures (with $\lambda^{-2}$ dust emissivity).
The tracks running off to the right hand side are pointing to the
locus of the ``pure" low temperature grey-body. \label{fig13}}
\end{figure}
\clearpage

\begin{figure}
\plotone{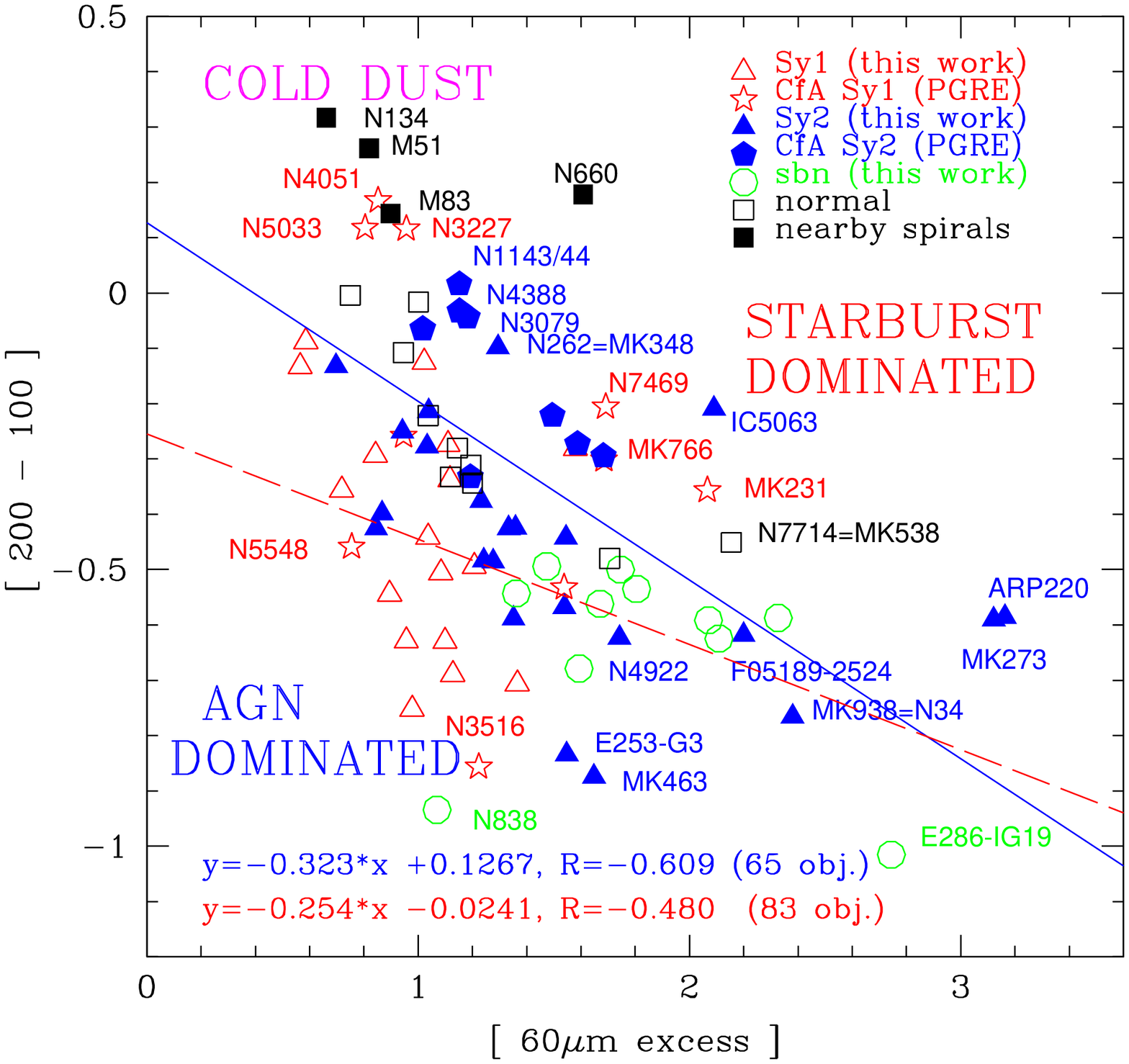} \caption{The [200 - 100] color versus the
60$\mu$m excess diagram of galaxies belonging to the 12$\mu$m
galaxy sample. The data for 17 CfA Seyfert galaxies (PGRE), 10
normal galaxies \citep{S99}, 4 nearby spiral galaxies \citep{A98}
and the Seyfert 2's NGC~7582 \citep{RA99}, all belonging to the
12$\mu$m galaxy sample, are included. The broken line shows the
linear fit to all the data of 12$\mu$m galaxies, while the solid
line shows the fit of all but the Seyfert 1's (the regression
coefficient is R=-0.480 for 83 data points and R=-0.609 for 65
points, respectively). The anti-correlation improves and becomes
more steep excluding Seyfert 1's. \label{fig14}}
\end{figure}
\clearpage

\begin{figure}
\plotone{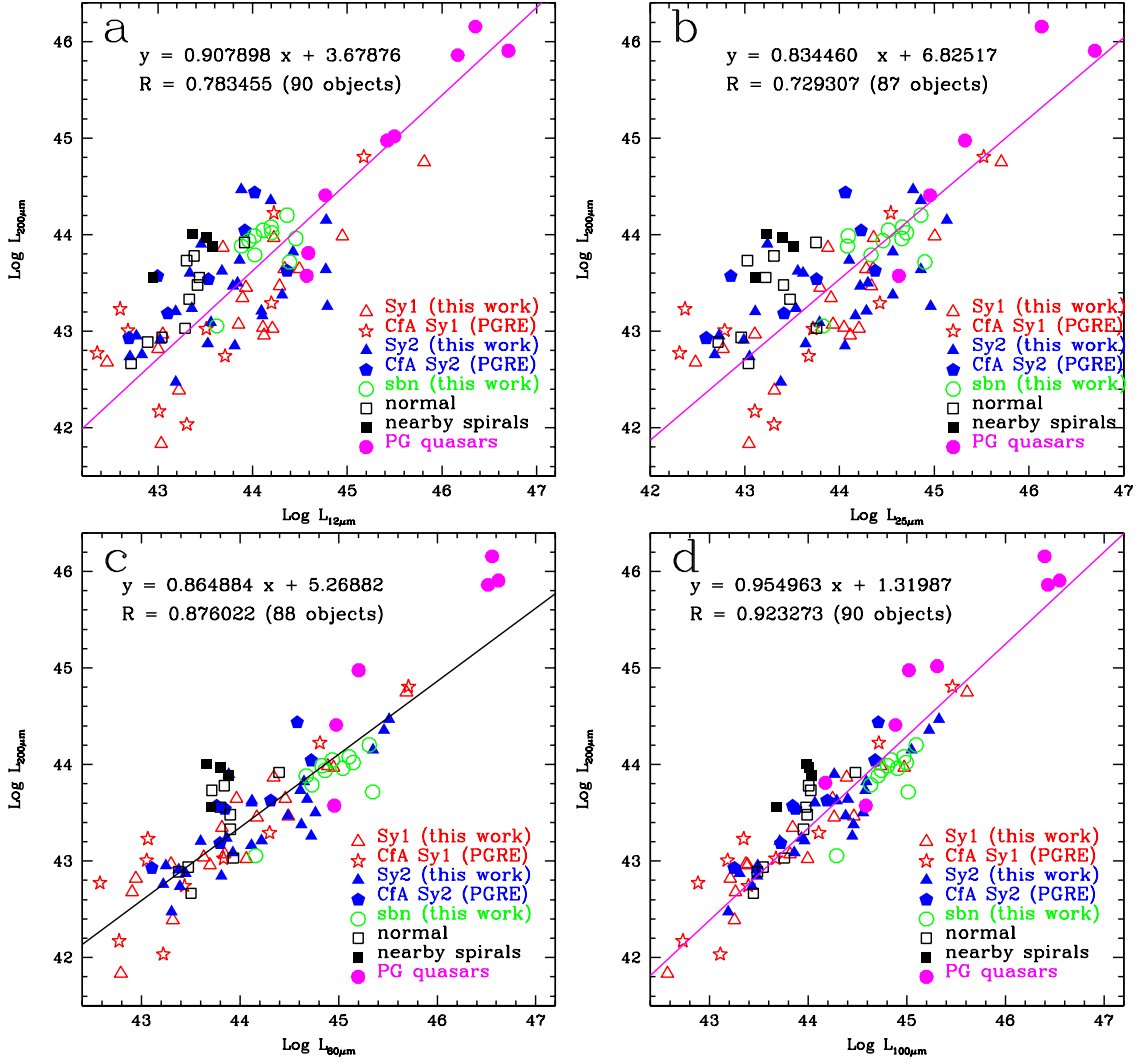} \caption{The 200 $\mu$m luminosity versus: (a)
the 12 $\mu$m luminosity; (b) the 25 $\mu$m luminosity; (c) the 60
$\mu$m luminosity; (d) the 100 $\mu$m luminosity. Note that the
luminosities (in this and in the following figures) are always in
units of erg~s$^{-1}$. \label{fig15}}
\end{figure}
\clearpage

\begin{figure}
\plotone{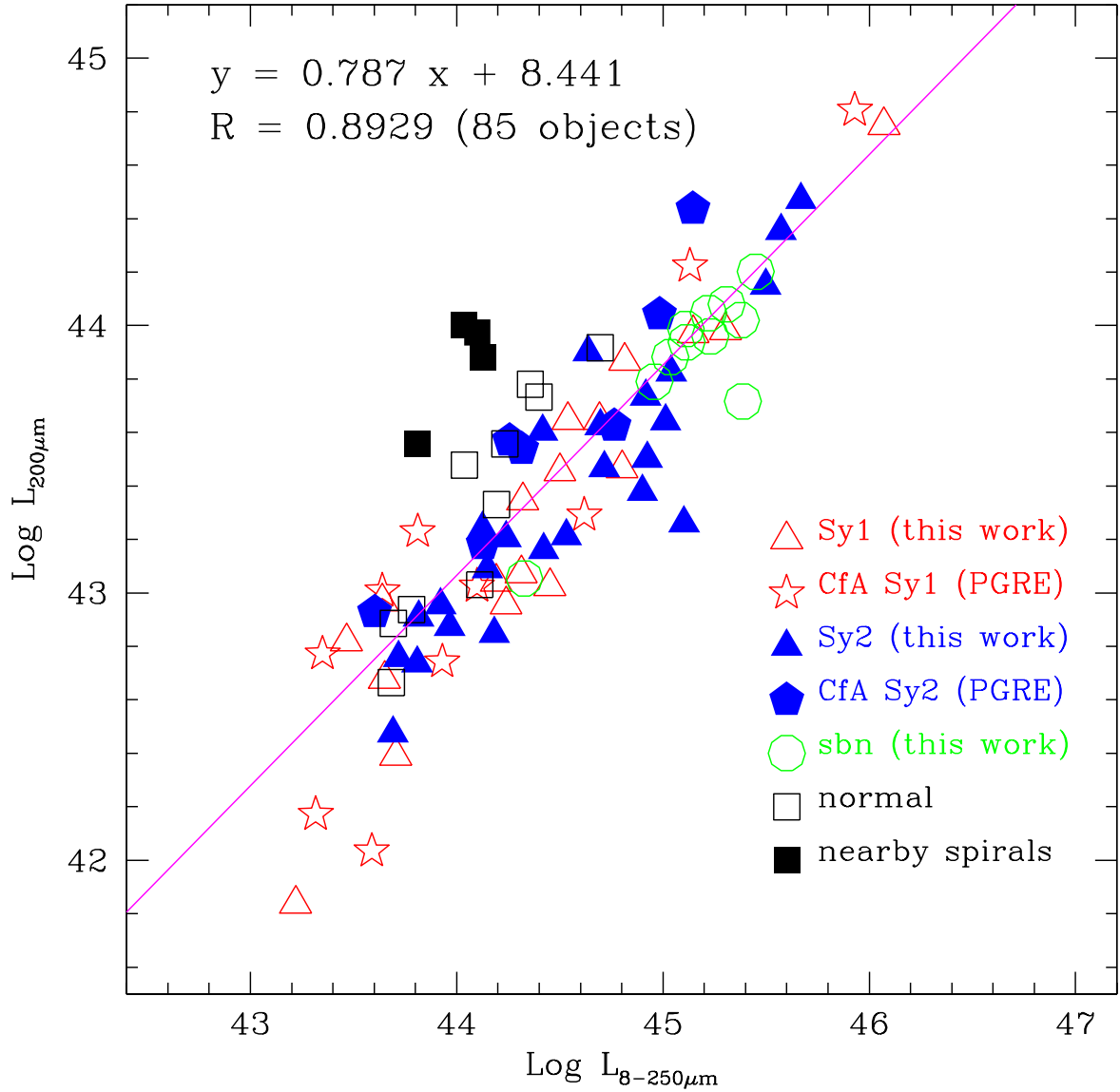} \caption{The 200 $\mu$m luminosity versus the
total mid to far-infrared $\mu$m luminosity (given in Table 3).
\label{fig16}}
\end{figure}
\clearpage

\begin{figure}
\plotone{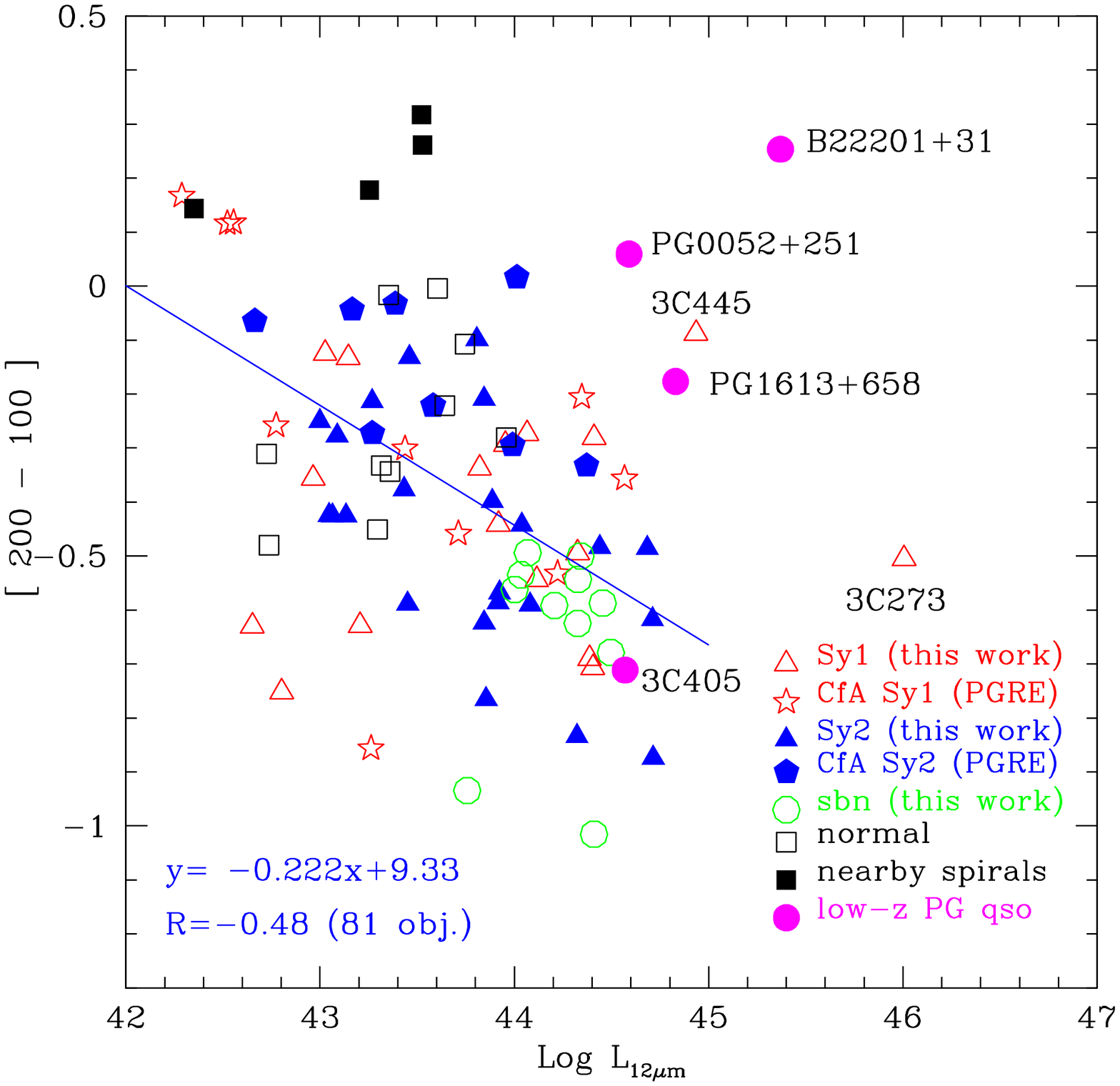} \caption{The [200 - 100] color versus the
12$\mu$m luminosity diagram of galaxies belonging to the 12$\mu$m
galaxy sample. For comparison are also shown three low redshift (z
$\leq$ 0.2) PG quasars and 3C405 \citep{H00}. The solid line
represents the linear fit that excludes the 3C objects and the PG
quasars, showing a mild correlation between color and 12$\mu$m
luminosity (the regression coefficient is R=-0.482 for 81 data
points). \label{fig17}}
\end{figure}
\clearpage

\begin{figure}
\plotone{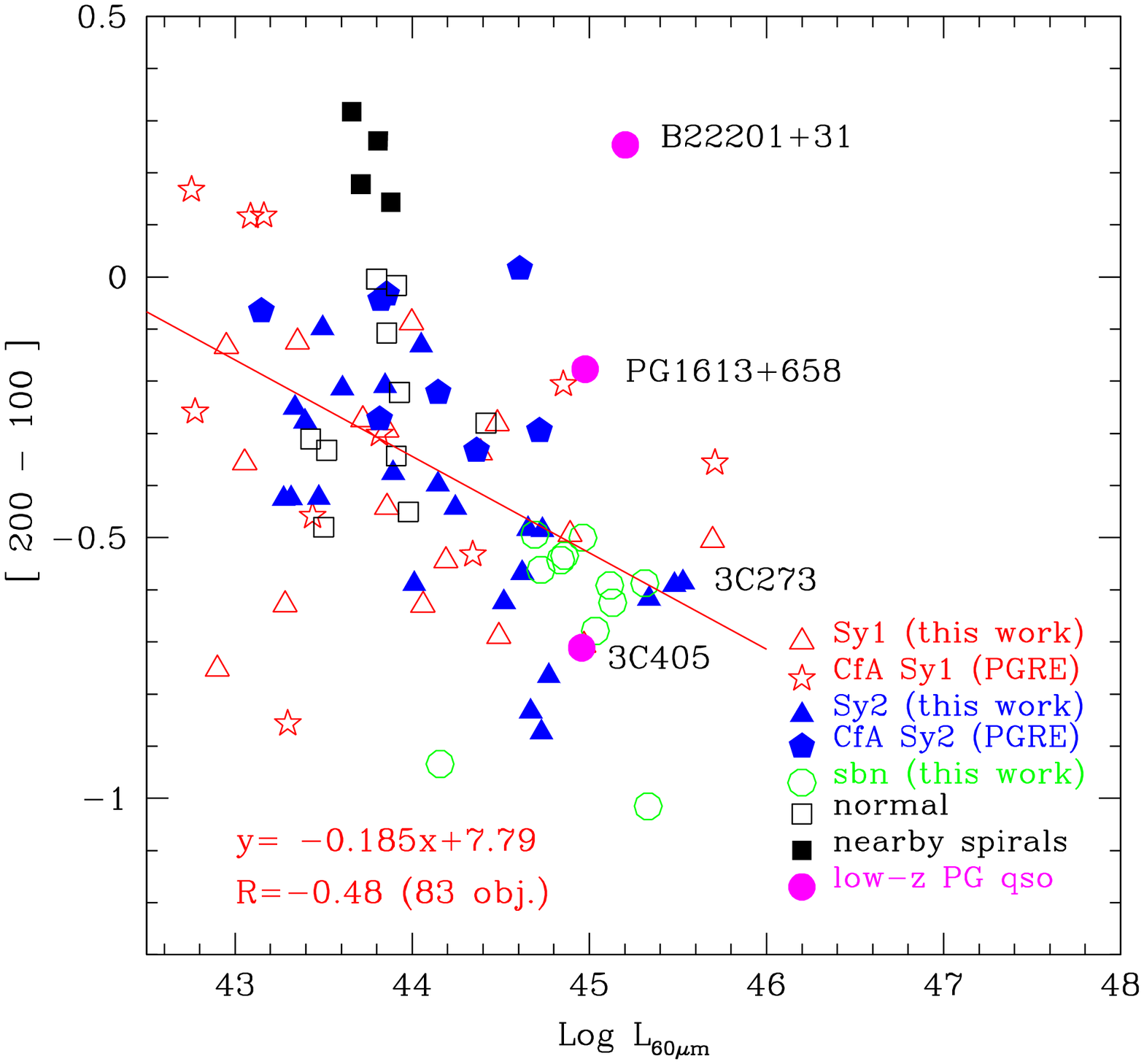} \caption{The [200 - 100] color versus the
60$\mu$m luminosity diagram of galaxies belonging to the 12$\mu$m
galaxy sample. For comparison are also shown two low redshift (z
$\leq$ 0.2) PG quasars and 3C405 \citep{H00}. The solid line
represents the linear fit that excludes the 3C objects and the PG
quasars, showing a mild correlation between color and 60$\mu$m
luminosity (the regression coefficient is R=-0.477 for 83 data
points). \label{fig18}}
\end{figure}
\clearpage

\begin{figure}
\plotone{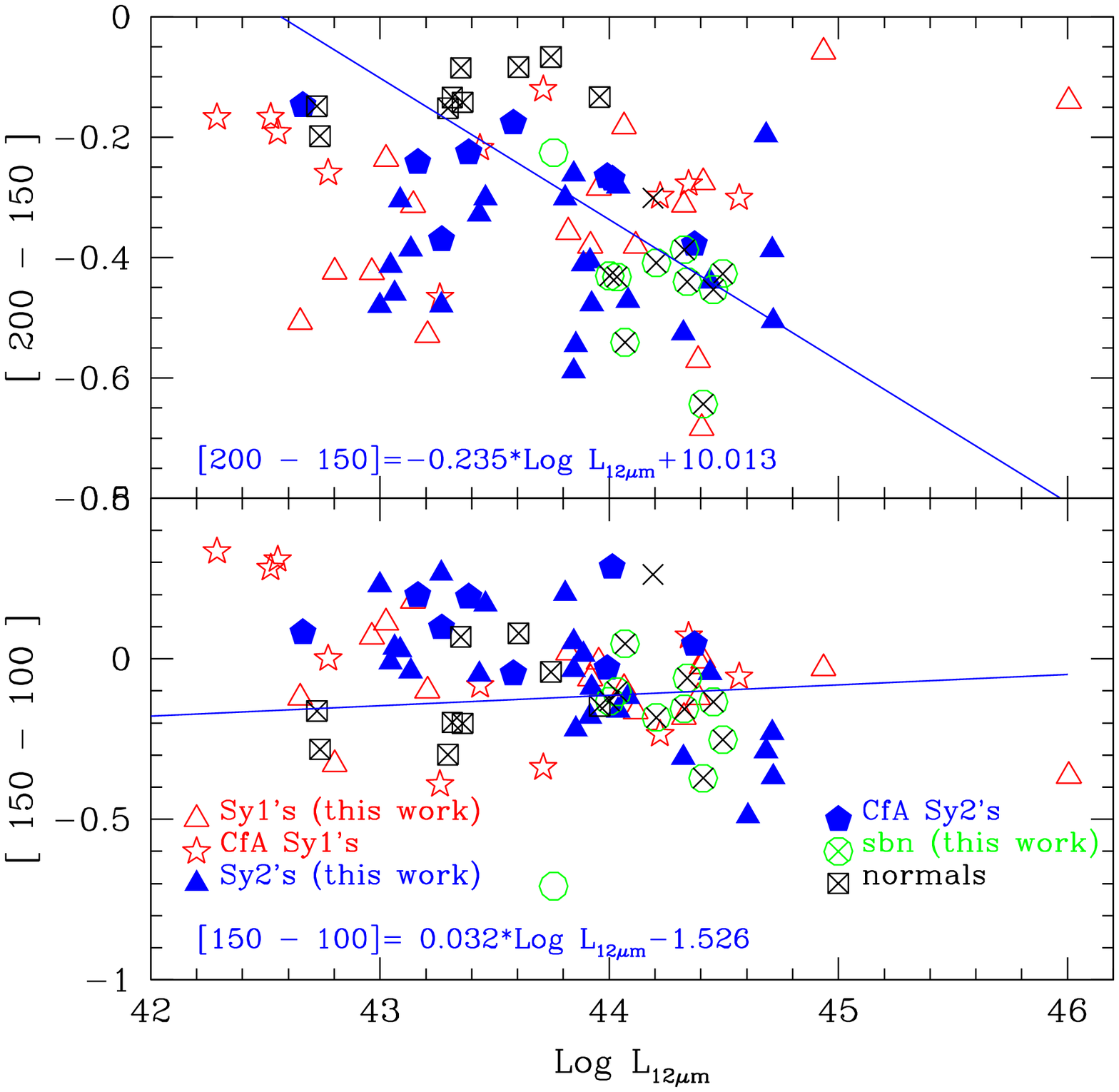} \caption{Intermediate colors [200 - 150] and
[150 - 100] versus 12$\mu$m luminosity. The solid line represents
the linear fit of the non-Seyfert galaxies (symbols marked with a
cross). \label{fig19}}
\end{figure}
\clearpage

\begin{figure}
\plotone{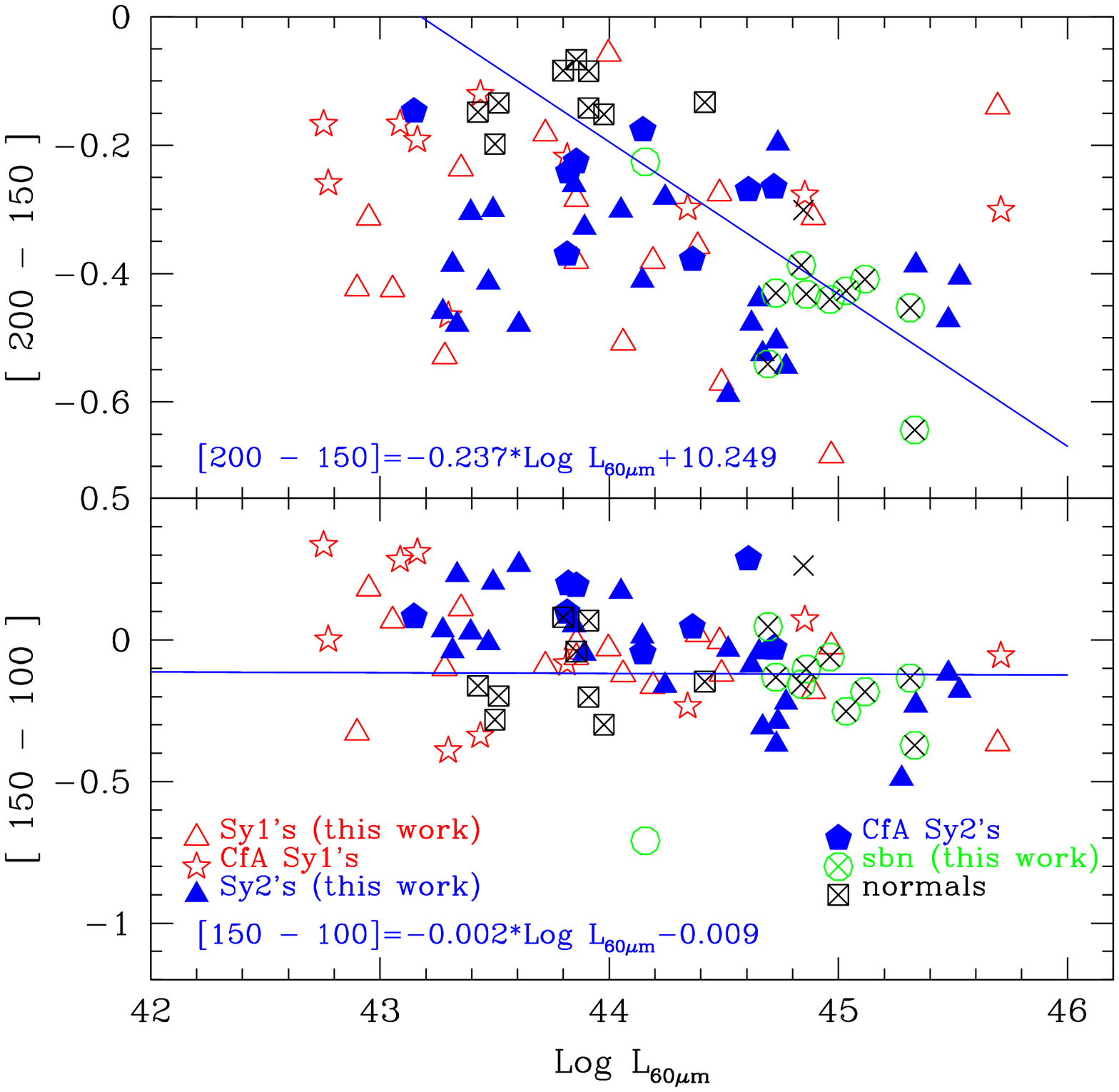} \caption{Intermediate colors [200 - 150] and
[150 - 100] versus 60$\mu$m luminosity. The solid line represents
the linear fit of the non-Seyfert galaxies (symbols marked with a
cross).\label{fig20}}
\end{figure}
\clearpage




\clearpage


\clearpage

\begin{deluxetable}{lrrrcccr}
\tabletypesize{\scriptsize} \tablecaption{Journal of the ISOPHOT
photometric observations of 12$\mu$m active galaxies
\label{tbl-1}} \tablewidth{0pt} \tablehead{ \colhead{Name} &
\colhead{R.A.(J2000.0)}   & \colhead{Dec.(J2000.0)}   & \colhead{
z } & \colhead{type\tablenotemark{a}}  & \colhead{Obs.id.} &
\colhead{Time} &
\colhead{ ref.}  \\
\colhead{} & \colhead{h~~~m~~~s~~~}  & \colhead{deg~~~'~~~"~~} &
\colhead{}  & \colhead{} &  \colhead{(TDT No.)}& \colhead{(sec)}
& \colhead{} }
 \startdata
E12-G21       &    0:40:47.3 & -79:14:20.0 & 0.0328 & sy1 & 55304297 & 1028 & (1)\\
IZW1          &    0:53:37.0 &  12:40:11.8 & 0.0604 & sy1 & 39502078 &  992 & (2)\\
MK1034        &    2:23:20.2 &  32:11:33.4 & 0.0380 & sy1 & 65600799 &  888 & (1)\\
N931=MK1040   &    2:28:14.4 &  31:18:42.1 & 0.0164 & sy1 & 65600803 & 1380 & (1)\\
NGC1365        &    3:33:36.5 & -36:08:23.1 & 0.0055 & sy1 & 80301415 &  230 & (1)\\
F03450+0055   &    3:47:40.2 &   1:05:13.8 & 0.0310  & sy1 & 79501331 &  974 & (2)\\
MK618         &    4:36:23.2 & -10:22:30.5 & 0.0347 & sy1 & 67901154 & 1302 & (1)\\
M-5-13-17     &    5:19:36.6 & -32:39:27.3 & 0.0125 & sy1 & 67802257 & 1062 & (1)\\
MK9           &    7:36:58.6 &  58:46:12.9 & 0.0399 & sy1 & 70501363 &  824 & (1)\\
MK79          &    7:42:32.0 &  49:48:35.9 & 0.0221 & sy1 & 70501666 &  818 & (1)\\
3C273         &   12:29:04.4 &  2:01:45.1  & 0.1579 & sy1/qso& 24100607 & 928  & (2)\\
N4593         &   12:39:39.4 & -05:20:39.3 & 0.0083  & sy1 & 24600427 &  424 & (2) \\
M-6-30-15     &   13:35:50.6 & -34:19:07.8 & 0.0077  & sy1 & 25600791 &  616 & (2) \\
F13349+2438   &   13:37:15.0 &  24:21:47.9 & 0.1070  & sy1 & 58000935 &  214 & (2)\\
I4329A        &   13:49:16.9 & -30:20:00.2 & 0.0161 & sy1 & 25702639 &  552 & (2)\\
F15091-2107   &   15:11:58.0 & -21:20:28.6 & 0.0446  & sy1 & 64400729 & 1082 & (2) \\
N7213         &   22:09:15.9 & -47:09:53.5 & 0.0059  & sy1 & 35401563 &  422 & (2) \\
3C445         &   22:23:47.5 &  -2:04:48.2 & 0.0562 & sy1 & 18701453 &  552 & (2)\\
N7314         &   22:35:45.8 & -26:02:59.9 & 0.0048 & sy1 & 53905190 &  448 & (1)\\
\tableline
MK938         &    0:11:06.6 & -12:06:26.7 & 0.0192 & sy2 & 55903371 &  444 & (1)\\
N262=MK348    &    0:48:47.1 &  31:57:25.0 & 0.0151 & sy2 & 59900759 & 2070 & (1)\\
F00521-7054   &    0:53:56.2 & -70:38:03.4 & 0.0700 & sy2 & 54201225  &  578 & (1)\\
M-2-8-39      &    3:00:30.4 & -11:24:56.1 & 0.0301 & sy2 & 81201040  & 1784 & (1)\\
N1241         &    3:11:16.2 &  -8:55:12.0 & 0.0072 & sy2 & 79601744 &  506 & (1)\\
N1320=MK607   &    3:24:48.7 &  -3:02:31.7 & 0.0099 & sy2 & 79501745 & 870 & (1)\\
F03362-1642   &    3:38:33.3 & -16:32:16.9 & 0.0360 & sy2 & 79601192 & 1512 & (1)\\
N1667         &    4:48:37.6 &  -6:19:11.8 & 0.0153 & sy2 & 82203176 &  464 & (1)\\
F05189-2524   &    5:21:01.4 & -25:21:45.0 & 0.0415  & sy2 & 86301930 &  196 & (2) \\
E253-G3       &    5:25:18.0 & -46:00:19.7 & 0.0407 & sy2 & 72501678 &  790 &(1)\\
N4501         &   12:31:59.5 &  14:25:16.8 & 0.0077 & sy2 & 23902627/23902607  &  474/142 & (1)\\
TOL1238-364   &   12:40:56.9 & -36:44:06.7 & 0.0109 & sy2 & 07800505 & 516 & (2) \\
N4922A/B      &   13:01:25.1 &  29:18:42.2 & 0.0237 & sy2 & 24500532/24500513 &  474/392 & (1)\\
M-3-34-64     &   13:22:24.4 & -16:43:42.2 & 0.0172  & sy2 & 25701422  & 1124 & (1)\\
MK273         &   13:44:42.1 &  55:53:12.8 & 0.0373 & sy2 & 14201730 & 180 & (2) \\
MK463         &   13:56:02.7 &   18:22:19.0 & 0.0505 & sy2 & 58001035 & 196 & (2) \\
ARP220        &   15:34:57.5 &  23:30:17.5 & 0.0182  & sy2/ulirg & 47600416 & 750 & (2)\\
F19254-7245   &   19:31:21.3 & -72:39:19.9 & 0.0615  & sy2 & 10100205 &  180 & (2) \\
N6810         &   19:43:33.9 & -58:39:20.7 & 0.0066 & sy2 & 29901651/29901642 &  474/160 & (1)\\
N6890         &   20:18:18.1 & -44:48:24.5 & 0.0081 & sy2 & 33602245 & 1762 & (1)\\
I5063         &   20:52:02.8 & -57:04:13.5 & 0.0113  & sy2 & 33301248 & 1124 & (1)\\
N7130=I5135   &   21:48:19.4 & -34:57:03.2 & 0.0162 & sy2 & 53904967 &  570 & (1)\\
N7172         &   22:02:02.1 & -31:52:08.8 & 0.0086 & sy2 & 53905081 &  460 & (1)\\
F22017+0319   &   22:04:19.2 &   3:33:50.2 & 0.0660 & sy2 & 54000168 & 2488 & (1)\\
N7496         &   23:09:46.5 & -43:25:43.1 & 0.0055 & sy2 & 54201073 &  586 & (1)\\
N7590         &   23:18:55.2 & -42:14:16.9 & 0.0050 & sy2 & 54200878 &  562 & (1)\\
N7674=MK533   &   23:27:56.7 &   8:46:44.7 & 0.0290 & sy2 & 55800996 & 834 & (1)\\
\tableline 
MK551         &    0:29:25.1 &  30:33:33.8 & 0.0500 & sbn & 58302885 & 1574 & (1)\\
N232          &    0:42:45.8 & -23:33:37.2 & 0.0222 & sbn & 57001381 &  450 & (1)\\
IC1623        &    1:07:46.4 & -17:30:26.7 & 0.0201 & sbn & 57001288 &  376 & (1)\\
N838          &    2:09:38.4 & -10:08:48.0 & 0.0128 & sbn & 80600207 &  244 & (1)\\
U2369         &    2:54:01.7 &  14:58:15.1 & 0.0307 & sbn & 79300510 & 982 & (1)\\
M+1-33-36     &   13:01:49.9 &   4:20:01.5 & 0.0375 & sbn & 62000121 &  644 & (1)\\
IZW107        &   15:18:06.3 &  42:44:36.5 & 0.0400 & sbn & 57100715 &  610 & (1)\\
MK496=N6090   &   16:11:40.7 &  52:27:29.3 & 0.0291 & sbn & 54100719 &  886 & (1)\\
N6240         &   16:52:58.8 &   2:24:09.6 & 0.0243 & sbn & 47600512 &  752 & (2) \\
U11284        &   18:33:37.2 &  59:53:22.2 & 0.0286  & sbn & 25103948 &  419 & (1)\\
F20193-2013   &   20:22:15.4 & -20:04:03.3 & 0.0645 & sbn & 54200426 & 1370 & (1)\\
E286-IG19     &   20:58:26.6 & -42:38:57.3 & 0.0427  & sbn & 13500535 &  180 & (2) \\
E148-IG02     &   23:15:46.8 & -59:03:14.6 & 0.0446 & sbn & 13501750 &  180 & (2) \\
 \enddata

\tablecomments{ (1): observations belonging to the ISO Open Time
Program ``IR energy distributions and imaging of the complete
sample of 12${\mu}m$ active galaxies"; (2): ISO archive data.}
\tablenotetext{a}{The galaxy type is coded as follows: sy1:
Seyfert 1; sy2: Seyfert 2; sbn: starburst nucleus; ulirg:
ultraluminous infrared galaxy. }

\end{deluxetable}

\clearpage


\begin{deluxetable}{lcccccc}
\tabletypesize{\scriptsize} \tablecaption{Measured fluxes from
ISOPHT C-200 obsrvations of 12$\mu$m active galaxies, with
1$\sigma$ uncertainties. \label{tbl-2}} \tablewidth{0pt}
\tablehead{ \colhead{Name} & \colhead{type}   &
\colhead{F(120${\mu} m$)}   & \colhead{F(150${\mu} m$)} &
\colhead{F(170${\mu} m$)} & \colhead{F(180${\mu} m$)} &
\colhead{F(200 ${\mu} m$)} \\
\colhead{} & \colhead{}  & \colhead{(Jy)} & \colhead{(Jy)}  &
\colhead{(Jy)} &  \colhead{(Jy)}& \colhead{(Jy)}} \startdata
E12-G21    & sy1 & \nodata          & 2.21 $\pm$ 0.07   & 1.81 $\pm$ 0.06 & 1.14 $\pm$  0.02   & 0.92 $\pm$ 0.03 \\
IZW1       & sy1 & 2.57$\pm$ 0.07   & 1.89 $\pm$ 0.06   & 1.60 $\pm$ 0.05 & 1.16$\pm$ 0.07     & 0.92 $\pm$ 0.05  \\
MK1034     & sy1 & \nodata          & 10.9 $\pm$ 0.1    & \nodata             & 5.68 $\pm$  0.05   & 2.26 $\pm$ 0.10 \\
N931=MK1040& sy1 & \nodata          & 5.54 $\pm$ 0.13   & 5.25 $\pm$0.07  & 3.72 $\pm$  0.08   & 2.88 $\pm$ 0.03 \\
N1365      & sy1 & 217. $\pm$ 0.8   & 194.0$\pm$ 0.4    & 167.0 $\pm$ 0.6 & 103.0$\pm$ 0.4     & 85.2 $\pm$ 0.5  \\
F03450+0055& sy1 & \nodata          & 1.03 $\pm$ 0.19   & \nodata             & \nodata        & 0.68 $\pm$ 0.15 \\
MK618      & sy1 & 3.32 $\pm$ 0.17  & 3.16 $\pm$ 1.50   & \nodata             & 1.19 $\pm$  0.07   & 0.85 $\pm$ 0.04 \\
M-5-13-17  & sy1 & 2.54 $\pm$ 0.05  & 1.86 $\pm$ 0.09   & \nodata             & 0.86 $\pm$  0.05   & 0.55 $\pm$ 0.03  \\
MK9        & sy1 & \nodata          & 0.74 $\pm$ 0.11   & \nodata             & 0.36 $\pm$  0.04   & 0.23 $\pm$ 0.04  \\
MK79       & sy1 & 2.74 $\pm$ 0.12  & 2.04 $\pm$ 0.27   & 1.88 $\pm$ 0.15 & 1.12 $\pm$  0.07   & 0.85 $\pm$ 0.04  \\
3C273      & sy1/qso& 1.49$\pm$ 0.13& 1.09 $\pm$ 0.08   & 1.10 $\pm$ 0.05 & 0.80$\pm$ 0.11     & 0.79 $\pm$ 0.11   \\
N4593      & sy1 & \nodata          & 8.1  $\pm$ 0.1    & \nodata             &  \nodata       & 4.7  $\pm$ 0.1  \\
M-6-30-15  & sy1 & \nodata          & 1.06 $\pm$ 0.04   & \nodata             &  \nodata       & 0.40 $\pm$ 0.06 \\
F13349+2438& sy1 & \nodata          & \nodata               & 0.35 $\pm$ 0.02 &  \nodata       & \nodata         \\
I4329A     & sy1 &   \nodata        & 1.87$\pm$ 0.08    &  \nodata            &  \nodata       & 1.2 $\pm$ 0.2  \\
F15091-2107& sy1 & \nodata          & 1.47$\pm$ 0.03    &  \nodata            &  \nodata       & 0.78 $\pm$ 0.07 \\
N7213      & sy1 & \nodata          & 13.62$\pm$0.35    &  \nodata            & \nodata        & 6.6 $\pm$ 0.4 \\
3C445      & sy1 &   \nodata        & 0.56$\pm$ 0.08    & \nodata             & \nodata        & 0.49 $\pm$  0.16  \\
N7314      & sy1 & 24.5 $\pm$ 0.5   & 19.4 $\pm$ 0.4    & 17.4 $\pm$ 0.3  & 11.3 $\pm$ 0.2     & 7.3  $\pm$ 0.2  \\
\tableline
MK938=N34  & sy2 & 17.2 $\pm$ 0.4   & 10.6 $\pm$0.5      & 8.5$\pm$0.2   & 5.4$\pm$0.1      & 3.0$\pm$0.1  \\
N262=MK348 & sy2 & 2.24 $\pm$ 0.08  & 2.28 $\pm$0.04     & 1.89$\pm$0.03   & 1.29$\pm$0.02      & 1.14$\pm$0.04 \\
F00521-7054& sy2 & \nodata          & 0.22$\pm$0.02   &  \nodata            & 0.087$\pm$0.017    &  \nodata          \\
M-2-8-39   & sy2 & 1.46 $\pm$ 0.05  & 1.26 $\pm$0.03     &   \nodata           & 0.762$\pm$0.015    &    \nodata        \\
N1241      & sy2 & 17.6 $\pm$ 0.6   & 18.2 $\pm$3.6      & 14.6$\pm$2.6    & 8.12$\pm$0.10      & 6.03$\pm$0.09  \\
N1320=MK607& sy2 & 3.7 $\pm$ 0.1  & 2.58 $\pm$0.36     &   \nodata           & 1.44$\pm$0.05      & 1.06$\pm$0.04  \\
F03362-1642& sy2 & 1.55 $\pm$ 0.14  & 0.59 $\pm$0.08     &   \nodata           & 0.39$\pm$0.03      & 0.39$\pm$0.03  \\
N1667      & sy2 & 29.6 $\pm$ 8.8   & 16.3 $\pm$0.5      & 17.0$\pm$2.8    & 9.03$\pm$0.15      & 6.33$\pm$0.17  \\
F05189-2524& sy2 & 10.6 $\pm$ 0.4  & 7.0 $\pm$0.4       &  \nodata            & 3.2$\pm$0.3      & 2.9$\pm$0.4 \\
E253-G3    & sy2 & 2.74 $\pm$ 0.17  & 1.71 $\pm$0.19     &   \nodata           & 0.86$\pm$0.14      & 0.51$\pm$0.04  \\
N4501      & sy2 & 116  $\pm$ 0.3  & 94.1 $\pm$0.9      &   \nodata           & 52.3$\pm$0.4       & 47.0$\pm$0.5   \\
TOL1238-364& sy2 &      \nodata     &    \nodata         & 7.42$\pm$0.11   & 4.60$\pm$0.06      & 3.62$\pm$0.11 \\
N4922A/B   & sy2 & 6.8$\pm$0.9      & 7.1 $\pm$1.8       &   \nodata           & 2.97$\pm$0.03      & 1.83$\pm$0.08 \\
M-3-34-64  & sy2 & 6.4$\pm$0.3      & 4.4 $\pm$0.4       &   \nodata           & 3.6$\pm$0.2        & 2.3$\pm$0.4 \\
MK273      & sy2 & 23.6$\pm$0.52    & 16.95$\pm$0.22     &   \nodata           & 7.48$\pm$0.20      & 5.75$\pm$0.30 \\
MK463      & sy2 & 2.26$\pm$0.40    & 0.80$\pm$0.15      &   \nodata           & $<$ 0.30           & $<$ 0.25      \\
ARP220     & sy2/ulirg & 117.3$\pm$0.2 & 79.4$\pm$0.2    & 71.4$\pm$0.1    & 41.1$\pm$0.1       & 31.2$\pm$0.3  \\
F19254-7245& sy2 & 3.5$\pm$0.5      & 2.6$\pm$0.3        &   \nodata          & 0.7$\pm$0.2     &  \nodata          \\
N6810      & sy2 & \nodata          & 29.8$\pm$0.1       &   \nodata         & 16.9$\pm$0.2       &  14.0$\pm$0.2  \\
N6890      & sy2 & 11.7$\pm$0.1    & 8.8$\pm$0.1      &   \nodata           & 5.09$\pm$0.03      & 4.36$\pm$0.05  \\
I5063      & sy2 & 4.97$\pm$0.35    & 4.13$\pm$0.35      &   \nodata           & 3.10$\pm$0.26      & 2.26$\pm$0.38 \\
N7130=IC5135& sy2& 28.9$\pm$1.7     & 21.9$\pm$3.0       & 16.9$\pm$1.0    & 10.8$\pm$0.6       & 7.29$\pm$0.47 \\
N7172      & sy2 & 32.0$\pm$9.2     & 22.9$\pm$4.7       & 19.0$\pm$3.2    & 12.6$\pm$0.4       & 7.60$\pm$0.43 \\
F22017+0319& sy2 & \nodata          & 0.85$\pm$0.13      &  \nodata      &    \nodata             & 0.54$\pm$0.02 \\
N7496      & sy2 & 20.8$\pm$0.1     & 16.2$\pm$0.2       & 13.9$\pm$0.5    & 8.89$\pm$0.12      & 6.25$\pm$0.07 \\
N7590      & sy2 & 25.9$\pm$0.1     & 22.8$\pm$2.7       & 17.6$\pm$0.2    & 12.4$\pm$0.8       & 7.91$\pm$0.08 \\
N7674=MK533& sy2 & 9.38$\pm$0.18    & 7.65$\pm$0.37      & 6.02$\pm$0.06   & 3.67$\pm$0.11      & 2.78$\pm$0.04  \\
\tableline 
MK551      & sbn & 5.11$\pm$0.10    & 3.42$\pm$0.10      & \nodata         & 1.68$\pm$0.03      & 1.28$\pm$0.03  \\
N232       & sbn & 18.1$\pm$0.4     & 18.9$\pm$5.2       & 15.7$\pm$3.9    & 9.11$\pm$1.5       & 5.44$\pm$0.09  \\
IC1623     & sbn & 36.5$\pm$0.3     & 26.7$\pm$0.3       & 23.5$\pm$0.4    & 12.80$\pm$0.07     & 9.69$\pm$0.09 \\
N838       & sbn & \nodata          & 4.07$\pm$0.13      &  \nodata        & 2.84$\pm$0.08      & 2.42$\pm$0.14 \\
U2369      & sbn & 11.1$\pm$0.1     & 8.71$\pm$0.22      &6.97$\pm$0.09    & 4.27$\pm$0.07      & 3.22$\pm$0.10 \\
M+1-33-36  & sbn & 7.25$\pm$0.06    & 5.90$\pm$0.08      & 6.48$\pm$1.50   & 3.45$\pm$0.04      & 2.42$\pm$0.05 \\
IZW107     & sbn & 9.84$\pm$0.09  & 11.3$\pm$4.7\tablenotemark{a}  & 9.0$\pm$3.8\tablenotemark{a} & 3.18$\pm$0.04     & 2.29$\pm$0.08 \\
MK496=N6090& sbn & 9.98$\pm$0.10    & 6.87$\pm$0.10      & \nodata         & 3.47$\pm$0.03      & 2.55$\pm$0.04 \\
N6240      & sbn & 23.7$\pm$0.3     & 18.2$\pm$0.2       & 16.7$\pm$0.1    & 9.5$\pm$0.1        & 7.1$\pm$0.3   \\
U11284     & sbn & 35.1$\pm$0.3     & 27.0$\pm$0.2       &  \nodata        & 14.5$\pm$0.2       & 13.5$\pm$0.3 \\
F20193-2013& sbn & 3.98$\pm$0.04    & 3.00$\pm$0.05      & \nodata         & 1.26$\pm$0.02      &  \nodata        \\
E286-IG19  & sbn & 7.0$\pm$ 0.3     &  4.4$\pm$ 0.2      & \nodata         & 2.1$\pm$0.1        & 1.0$\pm$0.3  \\
E148-IG02  & sbn & 11.4$\pm$0.6     & 7.95$\pm$0.25      & \nodata         & 4.31$\pm$0.18      & 2.8$\pm$0.3   \\
\enddata
\tablenotetext{a}{The low value of the signal to noise is due to
a large uncertainty in the internal calibrator measurements.}
\end{deluxetable}

\begin{deluxetable}{lcclcclcc}
\tabletypesize{\scriptsize} \tablecaption{Total integrated mid- to
far-infrared luminosities of 12$\mu$m active galaxies
\label{tbl-3}} \tablewidth{0pt} \tablehead{ \colhead{Name}
&\colhead{Log(L$_{FIR}$)} & \colhead{type}   & \colhead{Name}
&\colhead{Log(L$_{FIR}$)} & \colhead{type} &
\colhead{Name}&\colhead{Log(L$_{FIR}$)} & \colhead{type}   \\
\colhead{} & \colhead{(erg s$^{-1}$)}  & \colhead{} & \colhead{} &
\colhead{(erg s$^{-1}$)} &  \colhead{}& \colhead{} & \colhead{(erg
s$^{-1}$)}&  \colhead{} }
 \startdata
E12-G21                & 44.501 & sy1 & N262=MK348             & 43.966 &  sy2 &  N7496       &    43.808  &  sy2  \\
IZW1                   & 45.303 & sy1 & F00521-7054            & 45.005 &  sy2 &  N7590       &    43.718  &  sy2  \\
MK1034      & 45.147 & sy1 & N1068\tablenotemark{a} & 44.763 &  sy2 &  N7582\tablenotemark{b} &    44.416  &  sy2  \\
N931=MK1040            & 44.320 & sy1 & N1143/4\tablenotemark{a}& 45.145 & sy2 &  N7674=MK533 &    45.040  &  sy2  \\
N1365                  & 44.815 & sy1 & M-2-8-39               & 44.221 &  sy2 &  MK551       &    45.229  &  sbn  \\
F03450+0055            & 44.192 & sy1 & N1241                  & 43.922 &  sy2 &  N232        &    45.035  &  sbn  \\
MK618                  & 44.802 & sy1 & N1320=MK607            & 43.691 &  sy2 &  I1623       &    45.221  &  sbn  \\
M-5-13-17              & 43.705 & sy1 & F03362-1642            & 44.422 &  sy2 &  N838        &    44.328  &  sbn  \\
MK9                    & 44.452 & sy1 & N1667                  & 44.696 &  sy2 &  U2369       &    45.114  &  sbn  \\
MK79                   & 44.313 & sy1 & F05189-2524            & 45.498 &  sy2 &  M+1-33-36   &    45.111  &  sbn  \\
N3227\tablenotemark{a} & 43.640 & sy1 & E253-G3                & 44.900 &  sy2 &  IZW107      &    45.377  &  sbn  \\
N3516\tablenotemark{a} & 43.588 & sy1 & N3079\tablenotemark{a} & 44.256 &  sy2 &  MK496=N6090 &    44.958  &  sbn  \\
N4051\tablenotemark{a} & 43.350 & sy1 & N3982\tablenotemark{a} & 43.602 &  sy2 &  N6240       &    45.308  &  sbn  \\
N4151\tablenotemark{a} & 43.317 & sy1 & N4388\tablenotemark{a} & 44.315 &  sy2 &  F20193-2013 &    45.097  &  sbn  \\
MK766=N4253\tablenotemark{a} & 44.098 & sy1  & N4501           & 44.637 &  sy2 &  E286-IG19   &    45.387  &  sbn  \\
3C273                  & 46.071 & sy1 & TOL1238-364            & 44.149 &  sy2 &  E148-IG02   &    45.450  &  sbn  \\
N4593                  & 43.640 & sy1 & N4922A/B               & 44.715 &  sy2 &  N245=MK555\tablenotemark{c}  & 44.037  &  nor  \\
MK231\tablenotemark{a} & 45.930 & sy1 & M-3-34-64              & 44.532 &  sy2 &  U2936\tablenotemark{c}       & 44.357  &  nor  \\
N5033\tablenotemark{a} & 43.811 & sy1 & N5256=MK266\tablenotemark{a} & 44.983 & sy2 &  U2982\tablenotemark{c} & 44.696  &  nor  \\
M-6-30-15              & 43.219 & sy1 & MK273                  & 45.572 &  sy2 &  N5430=MK799\tablenotemark{c} & 44.194  &  nor  \\
F13349+2438            & 45.508 & sy1 & MK463                  & 45.102 &  sy2 &  N5719\tablenotemark{c}       & 43.692  &  nor  \\
I4329A                 & 44.240 & sy1 & N5929\tablenotemark{a} & 44.128 &  sy2 &  N6918\tablenotemark{c}       & 43.684  &  nor  \\
N5548\tablenotemark{a} & 43.930 & sy1 & ARP220                 & 45.669 &  sy2 &  N7083\tablenotemark{c}       & 44.398  &  nor  \\
MK817\tablenotemark{a} & 44.619 & sy1 & F19254-7245            & 45.415 &  sy2 &  N7624=MK323\tablenotemark{c} & 44.234  &  nor  \\
F15091-2107            & 44.693 & sy1 & N6810                  & 44.125 &  sy2 &  N7714=MK538\tablenotemark{c} & 44.111  &  nor  \\
N7213                  & 43.466 & sy1 & N6890                  & 43.815 &  sy2 &  N7798=MK332\tablenotemark{c} & 43.781  &  nor  \\
3C445                  & 44.540 & sy1 & I5063                  & 44.182 &  sy2 &  N134\tablenotemark{d}        & 44.034  &  nea  \\
N7314                  & 43.650 & sy1 & N7130=IC5135           & 44.916 &  sy2 &  N660\tablenotemark{d}        & 43.809  &  nea  \\
N7469\tablenotemark{a} & 45.131 & sy1 & N7172                  & 44.239 &  sy2 &  N5194=M51\tablenotemark{d}   & 44.100  &  nea  \\
MK938=N34              & 44.923 & sy2 & F22017+0319            & 45.013 &  sy2 &  N5236=M83\tablenotemark{d}   & 44.127  &  nea  \\
 \enddata
\tablenotetext{a}{ISOPHOT data from \citet{P-G00}}
\tablenotetext{b}{ISOPHOT data from \citet{RA99}}
\tablenotetext{c}{ISOPHOT data from \citet{S99}}
\tablenotetext{d}{ISOPHOT data from \citet{A98}}

\tablecomments{The galaxy type is coded as follows: sy1: Seyfert
1; sy2: Seyfert 2; sbn: starburst nucleus; nor: normal galaxy;
nea: nearby spiral galaxy. }
\end{deluxetable}

\begin{deluxetable}{lccccccc}
\tabletypesize{\scriptsize}

\tablecaption{Averaged colors with 1$\sigma$ uncertainties and
spectral indices for the various classes of galaxies belonging to
the 12$\mu$m galaxy sample . \label{tbl-4}}

\tablewidth{0pt}

\tablehead{ \colhead{Galaxy class} & \colhead{[60 - 25]} &
\colhead{[100 - 60]} & \colhead{[200 - 100]} &
\colhead{$\alpha_{25-60{\mu}m}$} & \colhead{$\alpha_{60-100
{\mu}m}$} & \colhead{$\alpha_{100-200{\mu}m}$} &
\colhead{$\alpha_{12-100{\mu}m}$} }
 \startdata
Seyfert 1's   & 0.40$\pm$0.30  & 0.23$\pm$0.15  & -0.43$\pm$0.21  & -1.1 & -1.0 & 1.4 & -1.0 \\
CfA Seyfert 1's        & 0.47$\pm$0.29  & 0.19$\pm$0.17  & -0.26$\pm$0.32  & \nodata  & \nodata  & \nodata & \nodata  \\
Seyfert 2's   & 0.62$\pm$0.29  & 0.16$\pm$0.18  & -0.47$\pm$0.21  & -1.6 & -0.7 & 1.6 & -1.4 \\
CfA Seyfert 2's        & 0.77$\pm$0.26  & 0.23$\pm$0.09  & -0.16$\pm$0.14  & \nodata  & \nodata  & \nodata & \nodata  \\
Starburst              & 0.82$\pm$0.16  & 0.11$\pm$0.09  & -0.66$\pm$0.21  & -2.1 & -0.5 & 2.2 & -1.7 \\
Normal galaxies        & 0.84$\pm$0.12  & 0.28$\pm$0.12  & -0.25$\pm$0.17  & -2.2 & -1.3 & 0.8 & -1.6 \\
Nearby spirals         & 0.83$\pm$0.10  & 0.39$\pm$0.15  &  0.22$\pm$0.08  & -2.2 & -1.7 & -0.7& -1.6 \\
\enddata
\end{deluxetable}

\clearpage

\begin{deluxetable}{lccccccl}

\tabletypesize{\scriptsize} \tablecaption{Measured fluxes from
ISOPHT C-100 observations of 12$\mu$m active galaxies, with
1$\sigma$ statistical uncertainty. \label{tbl-5}}
\tablewidth{0pt} \tablehead{ \colhead{Name}    &
\colhead{F(60$\mu m$)} & \colhead{F(65$\mu m$)} &
\colhead{F(80$\mu m$)}  &
\colhead{F(90$\mu m$)} & \colhead{F(100$\mu m$)} &\colhead{F(105$\mu m$)} & \colhead{notes}\\
\colhead{}  & \colhead{(Jy)} & \colhead{(Jy)}  & \colhead{(Jy)}
&  \colhead{(Jy)}& \colhead{(Jy)}& \colhead{(Jy)}& \colhead{}}
 \startdata
F00521-7054 & 1.26$\pm$0.04 & 1.10$\pm$0.08 & 1.10$\pm$0.04 & 0.63$\pm$0.03 & \nodata         & 0.83$\pm$0.04 & \\
N7674       & 3.5$\pm$0.1   & 4.6$\pm$0.1   & 4.0$\pm$0.1   & 2.99$\pm$0.06 & \nodata         & 3.7$\pm$0.1   & \\
F03362-1642 &  1.10$\pm$0.05 & \nodata           & 1.10$\pm$0.04 & 1.01$\pm$0.03 &\nodata          & 1.10$\pm$0.04 & \\
M+1-33-36   &  5.5$\pm$1.0   & \nodata           & 5.7$\pm$1.0   & \nodata           & 7.8$\pm$1.0 & 5.2$\pm$1.0   & (1) \\
N262        &  0.74$\pm$0.07 & 0.67$\pm$0.07 & 0.87$\pm$0.05 & 0.73$\pm$0.03 & 0.93$\pm$0.04 & \nodata         & (1) \\
N1365       &  81.9$\pm$1.0  & 121.0$\pm$1.0 & \nodata           & 114.0$\pm$1.0 &\nodata          & \nodata        &   \\
N4501       &  18.0$\pm$0.2  & \nodata           & \nodata           & 35.0$\pm$0.4  &\nodata          & \nodata    & (1) \\
N4922AB     &  5.20$\pm$0.1  & \nodata           & \nodata           & 6.10$\pm$0.1  &\nodata          & \nodata    &  \\
N6890       &  8.4$\pm$0.6   & \nodata           & \nodata           & 15.8$\pm$0.9  &\nodata          & \nodata    &  \\
\enddata
\tablecomments{(1): extended emission detected}
\end{deluxetable}


\begin{thebibliography}{}
\bibitem[Alexander \& Aussel(1999)]{A99}
Alexander D.M. and Aussel, H., 1999, in ISO Surveys of a Dusty
Universe, Eds. D. Lemke, M. Stickel \& K. Wilke , Springer Lecture
Notes of Physics Series, 548, 117.
\bibitem[Alton et al.(1998)]{A98}
Alton, P.B., et al.1998, \aap, 335, 807
\bibitem[Antonucci(1993)]{anto93}
Antonucci, R. 1993, \araa, 31, 473
\bibitem[Calzetti et al.(2000)]{CA00}
Calzetti, D., Armus, L., Bohlin, R.C.,Kinney, A.L., Koornneef, J.
Storchi-Bergmann, T. 2000, \apj, 533, 682
\bibitem[Edelson \& Malkan(1986)]{EM86}
Edelson, R.A. \& Malkan, M.A. 1986, \apj, 308, 59.
\bibitem[Edelson, Malkan \& Rieke (1987)]{EMR87}
Edelson, R.A., Malkan, M.A. \& Rieke, G.H. 1987, \apj, 321, 233
\bibitem[Fang et al.(1998)]{FA98}
Fang, F., Shupe, D.L., Xu, C., Hacking, P.B. 1998, \apj, 500, 693
\bibitem[Franceschini et al.(2001)]{FRA01}
Franceschini A., Aussel H., Cesarsky C.J., Elbaz, D., Fadda, D.
2001, \aap, 378, 1
\bibitem[Gabriel et al.(1997)]{G97}
Gabriel, C., Acosta-Pulido, J., Heinrichsen, I., Morris, H., Tai,
W.-M. 1997, in ASP Conf. Ser. 125, Astronomical Data Analysis
Software and Systems VI, ed. Gareth Hunt and H. E. Payne, eds.,
108
\bibitem[Genzel et al.(1998)]{Gen98}
Genzel, R. et al. 1998, \apj, 498, 579
\bibitem[Genzel \& Cesarsky(2000)]{GC00}
Genzel, R. \& Cesarsky, C.J. 2000, \araa, 38, 761
\bibitem[Granato \& Danese(1994)]{gra94}
Granato, G.L., \& Danese, L., 1994, \mnras, 268, 235
\bibitem[Haas et al.(2000)]{H00}
Haas, M., Muller, S.A.H., Chini, R., Meisenheimer, K., Klaas, U.,
Lemke, D., Kreysa, E., Camenzind, M. 2000, \aap, 354, 453
\bibitem[Huchra \& Burg(1992)]{HB92}
Huchra, J. \& Burg, R. 1992, \apj, 393, 90
\bibitem[Kessler et al.(1996)]{K96}
Kessler M.F., et al. 1996, \aap, 315, L27
\bibitem[Klaas et al.(1997)]{K97}
Klaas, U., Haas, M., Heinrichsen, I., Schulz, B. 1997, \aap 325,
L21
\bibitem[IRAS PSC(1988)]{iras88}
IRAS Point Source Catalog, Version 2, 1988, Joint IRAS Science
Working Group (Washington, DC: GPO).
\bibitem[Lemke et al.(1996)]{L96}
Lemke D. et al. 1996, \aap 315, L64
\bibitem[Laureijs et al.(2001)]{Lau01}
Laureijs R.J.  et al. 2001,
    The ISO Handbook, Vol.V, The Imaging Photo-polarimeter, SAI/99-069/Dc,
    Version 1.2, July 1, 2001, T. Muller \& J. Blommaert
    Eds.(available also at http://www.iso.vilspa.esa.es/manuals/HANDBOOK/V/pht\_hb/)
\bibitem[Maiolino et al.(1995)]{mai95}
Maiolino, R., Ruitz, M., Rieke, G.H., Keller, L.D. 1995, \apj,
446, 561
\bibitem[Malkan(2001)]{MA01}
Malkan, M. 2001, preprint astro-ph/0110357, from Kyoto Cosmology
Conference in honor of Prof. Tomita.
\bibitem[Malkan(2000)]{MA00}
Malkan, M. 2000, in ``Birth and Evolution of the Universe",
Proceedings of Fourth RESCEU International Symposium, eds. K.
Sato and M. Kawasaki (Universal Academy Press; Tokyo), p. 119
(also available as astro-ph/0005251).
\bibitem[Meisenheimer et al.(2001)]{mei01}
Meisenheimer, K., Haas, M., M\"{u}ller, S.A.H., Chini, R., Klaas,
U., Lemke, D. 2001, \aap, 372, 719
\bibitem[Moshir et al.(1992)]{mo92}
Moshir, M. et al. 1992, Explanatory Supplement to the IRAS Faint
Source Survey, version 2, JPL D-10015 8/92 (Pasadena: JPL).
\bibitem[P\'erez Garc\'{\i}a \& Rodr\'{\i}guez Espinosa(2000)]{P-G00}
P\'erez Garc\'{\i}a, A.M. \& Rodr\'{\i}guez Espinosa, J.M. 2000,
\apj, 557, 39 (PGRE)
\bibitem[Polletta \& Courvoisier(1999)]{PC99}
Polletta, M. \& Courvoisier, T.J.-L., 1999, \aap, 350, 765
\bibitem[Polletta et al.(2000)]{PO00}
Polletta, M., Courvoisier, T.J.-L.,Hooper, E.J., Wilkes, B.J.
2000, \aap, 362, 75
\bibitem[Radovich et al.(1999)]{RA99}
Radovich, M., Klaas, U., Acosta-Pulido, J., Lemke, D. 1999, \aap,
348, 705
\bibitem[Rowan-Robinson and Crawford(1989)]{RR89}
Rowan-Robinson, M., and Crawford, J. 1989, \mnras, 238, 523 (RRC)
\bibitem[Rush, Malkan \& Spinoglio(1993)]{RMS93}
Rush, B., Malkan, M.A. \& Spinoglio, L. 1993, \apjs, 89,1 (RMS)
\bibitem[Siebenmorgen, Krugel \& Chini(1999)]{S99}
Siebenmorgen, R., Krugel, E., Chini, R. 1999, \aap, 351, 495
\bibitem[Schmidt \& Green(1983)]{SG83}
Schmidt, M. \& Green, R.F. 1983, \apj, 269, 352
\bibitem[Spinoglio et al.(1995)]{S95}
Spinoglio, L., Malkan, M.A., Rush B., Carrasco, L., Recillas-Cruz,
E. 1995, \apj, 453, 616 (S95)
\bibitem[Spinoglio \& Malkan(1989)]{spi89}
Spinoglio, L. \& Malkan, M.A. 1989, \apj, 342, 83
\bibitem[Thean et al.(2000)]{th00}
Thean, A., Pedlar, A., Kukula, M.J., Baum, S.A., O'Dea, C.P.
2000, \mnras, 314, 573
\bibitem[Thean et al.(2001)]{th01}
Thean, A., Pedlar, A., Kukula, M.J., Baum, S.A., O'Dea, C.P. 2001
\mnras, 325, 737
\bibitem[Xu et al.(1998)]{XU98}
Xu, C., et al. 1998, \apj, 508, 576
\end{thebibliography}
\end{document}